\documentclass{mn2e}
\usepackage{graphicx}
\usepackage{mncite}

\title[The mass of V347 Pup]
{The component masses of the cataclysmic variable V347~Puppis}

\author[T. D. Thoroughgood et al.]
{T. D. Thoroughgood,$^{1}$\thanks{E-mail: Tim.Thoroughgood@shef.ac.uk} 
V. S. Dhillon,$^{1}$ 
D. Steeghs,$^{2}$
C. A. Watson,$^{1}$
\newauthor
D. A. H. Buckley,$^{3}$
S. P. Littlefair,$^{4}$
D. A. Smith,$^{1,5}$
M. Still,$^{6}$\thanks{Also Universities Space Research Association} 
\newauthor
K. J. van der Heyden,$^{3}$
B. Warner$^{7}$ 
\\ \\
$^{1}$Department of Physics and Astronomy, University of Sheffield, 
Sheffield, S3 7RH, UK \\
$^{2}$Harvard-Smithsonian Center for Astrophysics, 60 Garden Street, MS-67, 
Cambridge, MA 02138, USA\\
$^{3}$South African Astronomical Observatory, PO Box 9, Observatory 7935, Cape 
Town, South Africa\\
$^{4}$School of Physics, University of Exeter, Stocker Road, Exeter, EX4, UK\\
$^{5}$Winchester College, Winchester, SO23 9LX, UK\\
$^{6}$NASA/Goddard Space Flight Center, Code 662, Greenbelt, MD 20771, USA\\
$^{7}$Department of Astronomy, University of Cape Town, Private Bag, Rondebosch 
7700, South Africa\\
}

\date{\center{\Large Accepted for publication in the Monthly 
Notices of the Royal Astronomical Society \\ 
\vspace{.5cm} \today}} 

\begin{document}
\maketitle

\begin{abstract}
We present time--resolved spectroscopy and photometry of the 
double--lined eclipsing cataclysmic variable V347~Pup (= LB 1800). 
There is evidence of irradiation on the inner hemisphere of the secondary star, 
which we correct for using a model to give a secondary star radial velocity 
of $K_R$ = 198 $\pm$ 5 km\,s$^{-1}$. 
The rotational velocity of the secondary star in V347~Pup is found to be 
$v \sin i$ = 131 $\pm$ 5 km\,s$^{-1}$ and the system inclination is 
$i = 84.0^\circ\pm2.3^\circ$. From these parameters we obtain 
masses of $M_1$ = 0.63 $\pm$ 0.04 $M_\odot$ for the white dwarf primary and 
$M_2$ = 0.52 $\pm$ 0.06 $M_\odot$ for the M0.5V secondary star, giving a 
mass ratio of $q$ = 0.83 $\pm$ 0.05. 
On the basis of the component masses, and the spectral type and radius of 
the secondary star in V347~Pup, we find tentative evidence for an evolved companion. 
V347~Pup shows many of the characteristics of the SW~Sex stars, exhibiting 
single--peaked emission lines, high--velocity S--wave components and phase--offsets 
in the radial velocity curve. We find spiral arms in the accretion disc of 
V347~Pup and measure the disc radius to be close to the maximum allowed in a 
pressureless disc.
\end{abstract} 


\section{Introduction}
\label{sec:introduction}

Cataclysmic variables (CVs) are close binary stars consisting of a red--dwarf 
secondary transferring material onto a white dwarf primary via an accretion 
disc or magnetic accretion stream. V347~Pup is an example of 
a nova--like variable (NL), a class of CV with high mass transfer rates and 
no recorded nova or dwarf--nova type outbursts; see \scite{warner95a} for a 
comprehensive review of CVs.

A knowledge of the masses of the component stars in CVs is 
fundamental to our understanding of the origin, evolution and behaviour of 
these systems. Population synthesis models (e.g. \pcite{kolb01}) and the 
disrupted magnetic braking model of CV evolution (e.g. \pcite{spruit83}; 
\pcite{rappaport83}) can be observationally tested only if the number 
of reliably known CV masses increases. One of the most reliable ways 
to measure the masses of CVs is to use the radial velocity and the rotational 
broadening of the secondary star in eclipsing 
systems. The radial velocity of the disc emission lines is often an unreliable 
indicator of the white dwarf motion because of contamination from, for 
example, the bright spot. At present, reliable masses are 
known for only $\sim$20 CVs, partly due to the difficulties in measurement (see 
\pcite{smith98} for a review).

V347~Pup was identified spectroscopically as a NL by \scite{buckley90} 
from the presence of high--excitation emission lines. 
Even though V347~Pup emits at X-ray wavelengths 
(as the {\em Uhuru} X-ray source 4U 0608--49), the 
NL classification was favoured over a magnetic CV class on account of the 
negligible polarisation present. The study by \scite{buckley90} 
revealed a bright and deeply eclipsing system, 
with a spectroscopic and photometric orbital period of 5.57 hrs. 
Their measured system inclination and emission--line radial velocity curve, 
together with an empirical secondary star mass estimated from the orbital period, 
suggested a high primary mass close to the Chandrasekhar limit.

A multiwavelength study by \scite{mauche94} revealed an X-ray 
spectral energy distribution similar to many dwarf novae in outburst, 
with a likely origin in an extended emission region rather than the boundary layer. 
The UV emission lines appear to have a similar origin and, in a later paper 
by \scite{shlosman96}, their behaviour in eclipse was sucessfully modelled 
as disc light scattered in a rotating wind. 
The presence of an accretion disc in V347~Pup was confirmed by a rotational 
disturbance of the optical emission lines through primary eclipse 
(\pcite{mauche94}; \pcite{still98}). The latter authors found evidence for 
spiral arms and disc overflow accretion, and identified 
the low-excitation optical emission profiles as a 
composite of emission from the accretion disc and secondary star.

Secondary star absorption lines were found by \scite{diaz99}, 
who measured the system parameters of V347~Pup using the 
radial velocity semi--amplitudes of the primary and secondary stars. 
The radial velocity of the optical emission lines in V347~Pup varies 
widely in the literature, with published values 
of 134 $\pm$ 9 km\,s$^{-1}$ (\pcite{buckley90}), 122 $\pm$ 19 km\,s$^{-1}$ 
(\pcite{mauche94}), 156 $\pm$ 10 km\,s$^{-1}$, 125 $\pm$ 13 km\,s$^{-1}$ 
(\pcite{still98}) and 193 $\pm$ 16 km\,s$^{-1}$ (\pcite{diaz99}). 
The radial velocities of the UV emission 
lines published by \scite{mauche94} ranged between 220 and 370 km\,s$^{-1}$ with 
large phase shifts between spectroscopic conjuction and photometric mid--eclipse. 
This wide range in values, and the known unreliability of using disc emission 
lines in NLs to determine the motion of the white dwarf (e.g. \pcite{dhillon97b}), 
makes the determination of system parameters from the secondary star 
features alone highly desirable. In this paper, we derive the 
system parameters from the radial and rotational velocities of the secondary 
star in V347~Pup.

\section{Observations and Reduction}
\label{sec:observations}

During January and December 1998 and January 1999, we obtained optical 
spectra of V347~Pup using the Cassegrain spectrograph + SITe1 CCD chip on 
the SAAO 1.9-m telescope. Simultaneous photometry was 
available for most of the spectra using the SAAO 1.0-m telescope with 
the TEK8 CCD chip. See Table~\ref{tab:journal} and its caption for 
full details.

On the December 1998 run, we observed 17 spectral type templates ranging from 
G7V--M5.5V and telluric stars to remove atmospheric features. We 
observed flux standards on both the 1.9-m and 1.0-m telescopes on all nights. 

The spectra and images were reduced using standard procedures (e.g. 
\pcite{dhillon94}; \pcite{thoroughgood01}). The photometry data were 
corrected for the effects of atmospheric extinction by subtracting the 
magnitude of a nearby comparison star. 
The absolute photometry is accurate to approximately $\pm$0.5 mJy; the 
relative photometry to $\pm$0.01 mag. Comparison arc 
spectra were taken every $\sim$40 min in order to calibrate the wavelength scale 
and instrumental flexure. 
The arcs were fitted with fourth--order polynomials with an 
rms scatter of better than 0.04\AA. 
Where possible, slit losses were then corrected for by multiplying each 
V347~Pup spectrum by the ratio of the flux in the spectrum (over the whole 
spectral range) to the corresponding photometric flux. 

\begin{table*}
{\protect\small
\caption{Journal of observations. During Jan 1998, we used Grating 
No.\,4 to give a wavelength range of $\sim$4200--5060\AA\ ($\lambda_{cen}$ 
= 4610\AA) at 0.99-\AA$\:$ (64 km\,s$^{-1}$) resolution. 
Grating No.\,4 was again used on 28 Dec 1998 and 23 Jan 1999 
to give a wavelength range of $\sim$4900--5720\AA\ ($\lambda_{cen}$ 
= 5290\AA) at 0.95-\AA\ (54 km\,s$^{-1}$) resolution. 
On 25 and 27 Dec 1998, we used Grating No.\,5 to give a wavelength range of 
$\sim$5960--6725\AA\ ($\lambda_{cen}$ = 6330\AA) at 0.88-\AA$\:$ 
(42 km\,s$^{-1}$) resolution. 
Simultaneous photometry for the December 1998 spectra was recorded in the 
Johnson--Cousins $V$ and $R$ bands. Photometry was also available during 
the January 1998 run in the Str\"{o}mgren $b$ and $y$ filters. 
The seeing measured around 1.0 arcsec with photometric conditions on 25 and 27 
Dec 1998 and Jan 23 1999. On 28 Dec 1998, however, the seeing was poor and 
patchy high cloud was present. The seeing varied between 1.0--1.5 arcsec over 
the January 1998 run.
The epochs are calculated using the new ephemeris presented in this paper 
(equation~\ref{eqn:ephem}).}
\label{tab:journal}
\begin{tabular*}{0.95\textwidth}{lcccr@{.}lr@{.}lcccr@{.}lr@{.}l}

\hline
\vspace{-3mm}\\
\multicolumn{1}{c}{UT Date}
& \multicolumn{1}{c}{1.9-m} 
& \multicolumn{1}{c}{No. of} 
& \multicolumn{1}{c}{Exposure} 
& \multicolumn{2}{c}{Epoch} 
& \multicolumn{2}{c}{Epoch} 
& \multicolumn{1}{c}{1.0-m} 
& \multicolumn{1}{c}{No. of}
& \multicolumn{1}{c}{Exposure} 
& \multicolumn{2}{c}{Epoch}
& \multicolumn{2}{c}{Epoch}\\
\multicolumn{1}{c}{} 
& \multicolumn{1}{c}{$\lambda_{cen}$ (\AA)}
& \multicolumn{1}{c}{spectra} 
& \multicolumn{1}{c}{time (s)}
& \multicolumn{2}{c}{start} 
& \multicolumn{2}{c}{end}
& \multicolumn{1}{c}{filter}
& \multicolumn{1}{c}{images}
& \multicolumn{1}{c}{time (s)}
& \multicolumn{2}{c}{start} 
& \multicolumn{2}{c}{end}\\
\vspace{-3mm}\\
\hline
\vspace{-3mm}\\
1998 Jan 07 & 4610 & 113 & 200 & 17178&62 & 17179&91 & $b$ & 843 & 30 
& 17178&88 & 17179&93 \\ 
1998 Jan 08 & 4610 & 32 & 200 & 17182&88 & 17183&23 & $b$ & 105 & 30 
& 17183&83 & 17184&04 \\ 
1998 Jan 10 & 4610 & 117 & 200 & 17191&94 & 17192&94 & $y$ & 1379 & 30 
& 17191&61 & 17192&96 \\ 
1998 Jan 11 & 4610 & 129 & 200 & 17195&74 & 17197&25 & $b$ & 1217 & 30 
& 17195&81 & 17197&08 \\ 
1998 Jan 12 & 4610 & 116 & 200 & 17200&09 & 17201&44 & $b$ & 373 & 30 
& 17200&14 & 17200&52 \\ 
1998 Dec 25 & 6330 & 68 & 300 & 18696&47 & 18697&55 & $R$ & 962 & 30 
& 18696&32 & 18697&65 \\ 
1998 Dec 27 & 6330 & 61 & 300 & 18705&18 & 18706&17 & $R$ & 748 & 30 
& 18704&95 & 18706&25 \\
1998 Dec 28 & 5290 & 23 & 300 & 18709&63 & 18710&16 & $V$ & 442 & 30 
& 18709&50 & 18710&29 \\
1999 Jan 23 & 5290 & 64 & 300 & 18821&30 & 18822&49 & \multicolumn{7}{c}
{no photometry available} \\
\vspace{-3mm}\\
\hline
\end{tabular*}
}
\end{table*}

\section{Results}
\subsection{Ephemeris}
\label{sec:ephem}

The times of mid--eclipse for V347~Pup were determined by 
fitting a parabola to the eclipse minima in the photometry data. 
A least--squares fit to the 21 eclipse timings listed in 
Table~\ref{tab:o-cs} yields the ephemeris:
\begin{equation}
\label{eqn:ephem}
\begin{array}{lr@{.}llr@{.}ll}
T_{\rm mid-eclipse} = & \!\!\!\! {\rm HJD}\,\,2\,446\,836&96176 
& \!\!\!\! + \!\!\!\! & 0&231936060 & \!\!\!\!\! E \\
& \!\! \pm \,\, 0&00009 & \!\!\!\! \pm \!\!\!\! & 
0&000000006. & \\
\end{array}
\end{equation}
Our new ephemeris is exactly the same as that given by \scite{baptista91}, 
except we 
have reduced the errors on both the zero point and orbital period.
We find no evidence for any systematic variation in the O--C values 
listed in Table~\ref{tab:o-cs}.

% Mid-eclipse timings for V347~Pup
\begin{table}
{\protect\small
\caption{Times of mid--eclipse for V347~Pup according to 
Buckley et al. (1990; B90), Baptista \& Cieslinski (1991; BC91) and this 
paper.} 
\label{tab:o-cs}
\begin{tabular*}{87mm}{lr@{.}lr@{x}lr@{.}lc}

\hline
\vspace{-3mm}\\
\multicolumn{1}{c}{Cycle} & 
\multicolumn{2}{c}{HJD} & \multicolumn{2}{c}{Uncertainty}
& \multicolumn{2}{c}{O--C} & 
\multicolumn{1}{c}{Reference} \\
\multicolumn{1}{c}{(E)} & 
\multicolumn{2}{c}{at mid--eclipse} & \multicolumn{2}{c}{on HJD}
& \multicolumn{2}{c}{(secs)} & \multicolumn{1}{c}{}\\
\multicolumn{1}{c}{} & 
\multicolumn{2}{c}{(2,400,000+)} & \multicolumn{2}{c}{}
& \multicolumn{2}{c}{} & 
\multicolumn{1}{c}{} \\
\vspace{-3mm}\\
\hline
\vspace{-3mm}\\
--4 & 46836&0379 & 5&$10^{-4}$ & 335&96 & B90\\
0 & 46836&9621 & 5&$10^{-4}$ & 29&74 & B90\\
39 & 46846&0059 & 5&$10^{-4}$ & --117&69 & B90\\
43 & 46846&9333 & 5&$10^{-4}$ & --147&43 & B90\\
48 & 46848&0930 & 5&$10^{-4}$ & --145&73 & B90\\
56 & 46849&9500 & 5&$10^{-4}$ & --15&13 & B90\\
65 & 46852&0373 & 5&$10^{-4}$ & --25&89 & B90\\
69 & 46852&9651 & 5&$10^{-4}$ & --21&08 & B90\\
78 & 46855&0533 & 5&$10^{-4}$ & 45&92 & B90\\
6177 & 48269&63136 & 1.5&$10^{-4}$ & 48&28 & BC91\\
7583 & 48595&73325 & 1.1&$10^{-4}$ & 30&04 & BC91\\
7587 & 48596&66022 & 9&$10^{-5}$ & --36&85 & BC91\\
17179 & 50821&39199 & 5&$10^{-4}$ & 56&29 & This paper\\
17184 & 50822&55127 & 5&$10^{-4}$ & 21&70 & This paper\\
17192 & 50824&40676 & 5&$10^{-4}$ & 21&83 & This paper\\
17196 & 50825&33428 & 5&$10^{-4}$ & 2&46 & This paper\\
17197 & 50825&56611 & 5&$10^{-4}$ & --6&70 & This paper\\
18697 & 51173&47035 & 1&$10^{-4}$ & 6&20 & This paper\\
18705 & 51175&32562 & 1&$10^{-4}$ & --12&68 & This paper\\
18706 & 51175&55802 & 1&$10^{-4}$ & 26&54 & This paper\\
18710 & 51176&48519 & 1&$10^{-4}$ & --22&21 & This paper\\
\vspace{-3mm}\\
\hline

\end{tabular*}
}
\end{table}

\subsection{Average spectrum}
\label{sec:average}

\begin{figure*}
\begin{center}
\includegraphics[width=8cm,angle=-90]{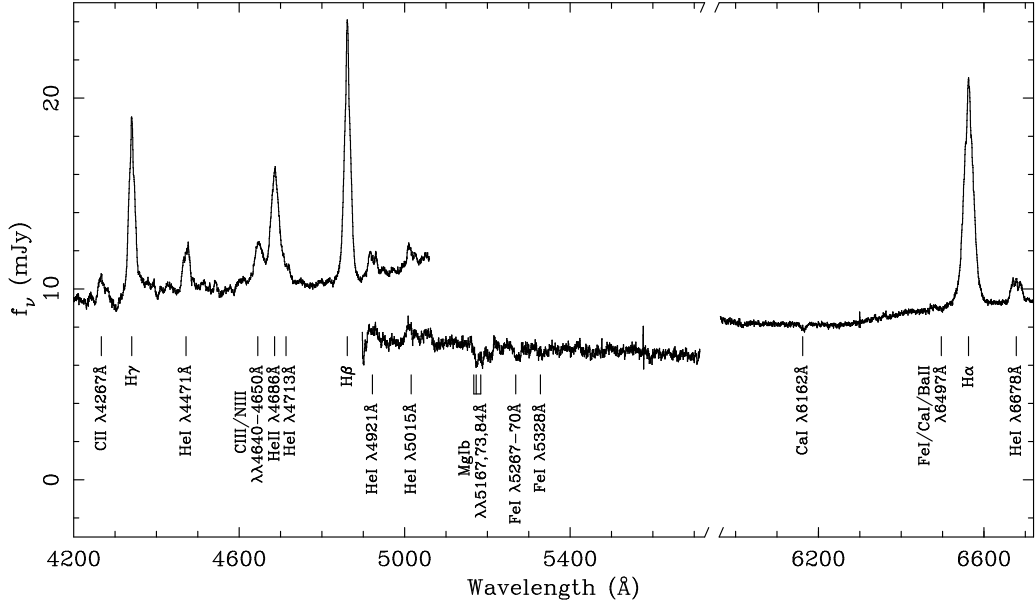}
\end{center}
\caption{\protect\small 
The average spectra for the three wavelength regions; the spectrum centred on 
$\lambda$4610\AA\ is an average of all spectra recorded on the Jan 1998 run, and has 
not been corrected for slit--losses. The spectrum centred on $\lambda$5290\AA\ is an 
average of all data recorded on 28 Dec 1998 and 23 Jan 1999, placed on an absolute 
flux scale (as determined from the 28 Dec 1998 photometry and flux standards). 
The spectrum centred on $\lambda$6330\AA\ is composed of all data from 25 and 
27 Dec 1998 and has been placed on an absolute flux scale. All average spectra 
are uncorrected for orbital motion, resulting in smeared spectral features.}
\label{fig:av_spec}
\end{figure*}

The average spectra of V347~Pup, uncorrected for orbital motion, are shown in 
Fig.~\ref{fig:av_spec}. In Table~\ref{tab:linewidths}, we list fluxes, 
equivalent widths and velocity widths of the most prominent 
lines measured from the average spectra. 

The Balmer emission lines are broad, symmetric and single--peaked, 
instead of the double--peaked profile one would expect from a high inclination 
accretion disc (e.g. \pcite{horne86}). This behaviour is characteristic 
of the SW~Sex stars (e.g. \pcite{dhillon97b}). 
Previous studies of V347~Pup by \scite{buckley90} and \scite{diaz99} agree 
with this single--peaked observation, however, the study by \scite{still98} 
shows double--peaked low excitation lines (although this could be 
due to the presence of absorption cores). 
The HeI $\lambda$6678\AA\ line appears to be composed of a narrow 
single--peaked component superimposed upon a broad double--peaked component. 
The other HeI emission lines can clearly be seen in the wavelength region 
centred on $\lambda$4610\AA\ as double peaked in nature, with the 
possible exception of HeI $\lambda$4471\AA. High--excitation 
emission is present through HeI $\lambda$4686\AA\ and the CIII/NIII 
$\lambda\lambda$4640--4650\AA\ Bowen fluoresence complex.

The secondary 
star is clearly visible in the average spectra as absorption 
lines of the neutral metals CaI, FeI and MgI, as seen in \scite{diaz99}. 
Secondary star features in the SW~Sex stars are not unusual in the longer 
period systems, such as BT~Mon (\pcite{sad98b}), AC~Cnc and V363~Aur 
(\pcite{thoroughgood04}).

\begin{table*}
{\protect\small
\caption{Fluxes and widths of prominent lines in V347~Pup, measured 
from the 2 nights' data centred on $\lambda$6330\AA\ and the night of 
11 Jan 1998 centred on $\lambda$4610\AA. The full--width half--maximum (FWHM) 
velocities were determined from Gaussian fits, whereas the full--width 
zero--intensity (FWZI) velocities and their errors have been estimated by eye. 
HeII$\:\lambda$4686\AA, CIII/NIII$\:\lambda\lambda$4640--4650\AA$\:$ 
and HeI$\:\lambda$4713\AA$\:$ are blended, so separate values 
of the flux and EW are given (determined from a triple--Gaussian fit) as 
well as the combined flux of the three.}
\label{tab:linewidths}
\begin{center}
\begin{tabular*}{0.772\textwidth}
{lcr@{$\:\pm\:$}lr@{$\:\pm\:$}lr@{$\:\pm\:$}lr@{$\:\pm\:$}l}
\hline
\vspace{-3mm}\\
\multicolumn{1}{l}{Line} & 
\multicolumn{1}{l}{Date} & 
\multicolumn{2}{c}{Flux} & 
\multicolumn{2}{c}{EW} & 
\multicolumn{2}{c}{FWHM} &
\multicolumn{2}{c}{FWZI}\\
\multicolumn{1}{c}{} & 
\multicolumn{1}{c}{} & 
\multicolumn{2}{c}{($\times$ 10$^{-14}$}
& \multicolumn{2}{c}{(\AA)} & \multicolumn{2}{c}{(km\,s$^{-1}$)} 
& \multicolumn{2}{c}{(km\,s$^{-1}$)}\\
\multicolumn{1}{c}{} & 
\multicolumn{1}{c}{} & 
\multicolumn{2}{c}{(ergs\,cm$^{-2}$\,s$^{-1}$)}
& \multicolumn{2}{c}{} & \multicolumn{2}{c}{} 
& \multicolumn{2}{c}{}\\
\vspace{-3mm}\\
\hline
\vspace{-3mm}\\
H$\alpha$ & 25 Dec 1998 & 16.80&0.04 & 35.1&0.1 & 1100&100 & 3500&300 \\
H$\alpha$ & 27 Dec 1998 & 27.98&0.05 & 36.4&0.1 & 1100&100 & 3600&300 \\
H$\beta$ & 11 Jan 1998 & 42.2&0.1 & 24.6&0.2 & 1000&100 & 2800&300 \\
H$\gamma$ & 11 Jan 1998 & 34.6&0.2 & 16.9&0.3 & 1100&100 & 2600&800 \\
HeI$\:\lambda$4471\AA & 11 Jan 1998 & 8.0&0.1 & 3.8&0.2 & 1150&100 & 1850&200 \\
HeI$\:\lambda$4921\AA & 11 Jan 1998 & 4.23&0.08 & 2.6&0.1 & 1300&100 & 2000&200 \\
HeI$\:\lambda$5015\AA & 11 Jan 1998 & 3.3&0.1 & 2.2&0.2 & 1250&100 & 2000&200 \\
HeI$\:\lambda$6678\AA & 25 Dec 1998 & 1.49&0.02 & 2.92&0.08 & 1250&100 & 1900&200 \\
HeI$\:\lambda$6678\AA & 27 Dec 1998 & 2.40&0.04 & 3.00&0.08 & 1300&100 & 1900&200 \\
CII$\:\lambda$4267\AA & 11 Jan 1998 & 3.8&0.1 & 1.7&0.4 & 900&200 & 1800&600 \\
HeII$\:\lambda$4686\AA & 11 Jan 1998 & 26.9&0.3 & 13.1&0.3 & 1450&150 & 
\multicolumn{2}{c}{} \\
CIII/NIII$\:\lambda\lambda$4640--4650\AA & 11 Jan 1998 & 11.7&0.1 
& 6.4&0.2 & 1700&150 & \multicolumn{2}{c}{} \\
HeI$\:\lambda$4713\AA & 11 Jan 1998 & 4.8&0.3 & 2.3&0.3 & 1500&300 & 
\multicolumn{2}{c}{} \\
HeII + CIII/NIII + & 11 Jan 1998 & 45.5&0.2 & 22.4&0.2 
& \multicolumn{2}{c}{} & \multicolumn{2}{c}{} \\
HeI$\:\lambda$4713\AA & \multicolumn{9}{c}{} \\
\hline\\
\end{tabular*}
\end{center}
}
\end{table*}

\subsection{Light curves}
\label{sec:light}

\begin{figure*}
\begin{center}
\begin{tabular}{c}
\multicolumn{1}{l}{(a)}\\
\includegraphics[width=17cm,angle=0]{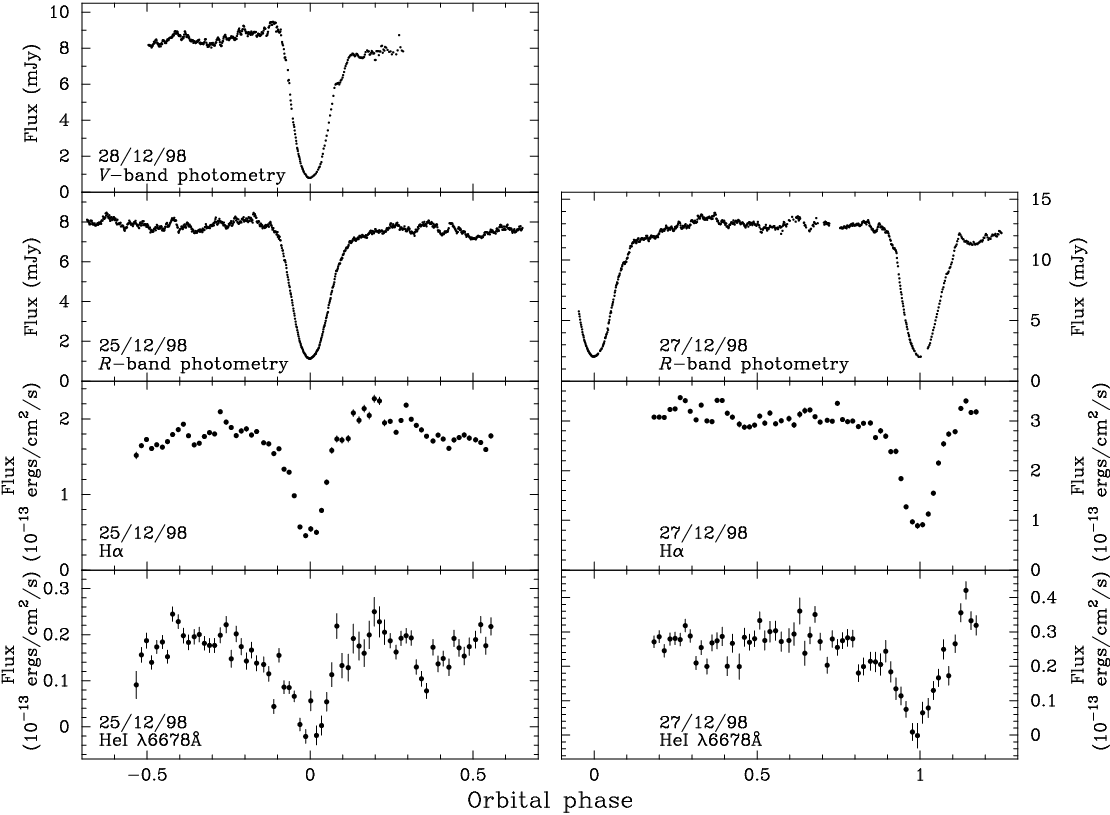}\\
\multicolumn{1}{l}{(b)}\\
\includegraphics[width=17cm,angle=0]{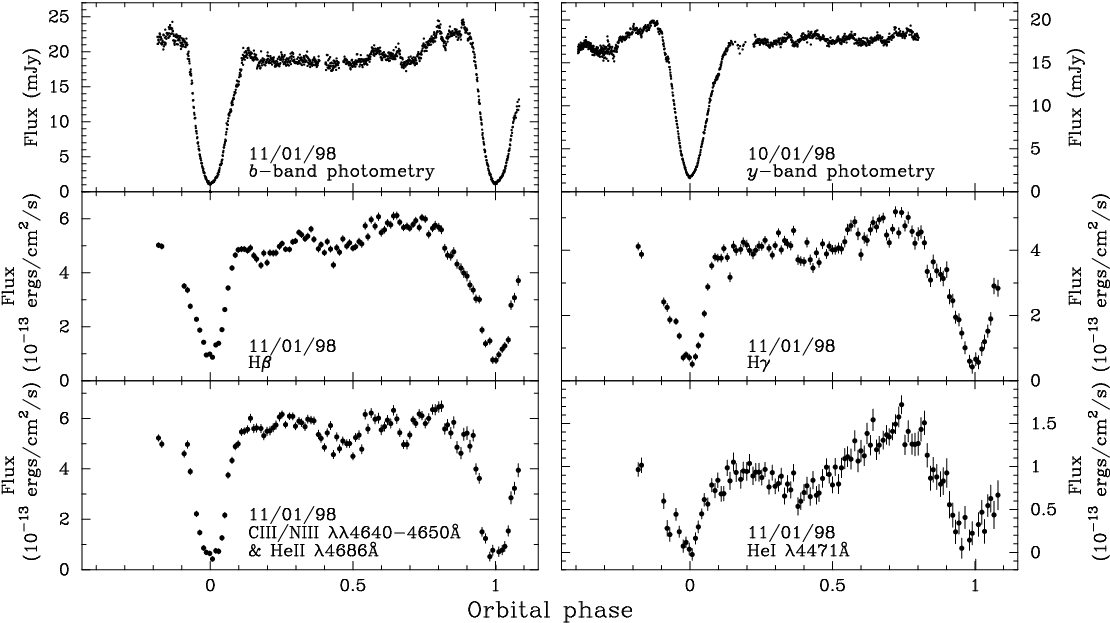}\\
\end{tabular}
\end{center}
\caption{\protect\small
Broad--band and emission--line light curves of V347~Pup recorded in 
Dec 1998 and Jan 1999 (a), and Jan 1998 (b); see the panel labels for details. 
Note the increase in continuum and emission--line flux 
between Dec 25 and Dec 27 1998.}
\label{fig:lc}
\end{figure*}

Fig.~\ref{fig:lc} shows the broad--band and emission--line light curves 
of V347~Pup. The emission--line light curves were produced by subtracting 
a polynomial fit to the continuum and summing the residual flux. All 
light curves are plotted as a function of phase following 
the ephemeris derived in Section~\ref{sec:ephem}. 

The $b$, $y$, $V$ and $R$--band light curves show deep, asymmetrical 
primary eclipses with the egress lasting longer than ingress. 
Flickering is present in all light curves, as well as an 
increase in brightness approaching eclipse in the $b$, $y$ and $V$--bands. 
The $b$ and $y$--band data recorded in 1998 Jan show no significant brightness 
variations during the run, with out-of-eclipse magnitudes of 13.3 $\pm$ 0.1 
in both filters. The eclipse depths are 3.2 and 2.6 mag, respectively. 
We measure $R$--band out-of-eclipse magnitudes of 14.00 $\pm$ 0.05 mag on 
Dec 25, increasing in brightness to 13.45 $\pm$ 0.10 mag on Dec 27. 
The eclipse depth remains roughly the same at 2.1 mag and 2.0 mag, 
respectively. In the $V$--band, the out-of-eclipse magnitude is 
14.10 $\pm$ 0.10 mag, with an eclipse depth of 2.6 mag. 
Photometric out-of-eclipse magnitudes 
in the literature range between 13.05--13.28 in $R$ and 13.2--13.58 in $V$ 
(\pcite{buckley90}, \pcite{mauche94}), suggesting that our observations in 
Dec 1998 find V347~Pup around 0.5--1 mag fainter. Long--term 
variations in the magnitudes of NLs are not uncommon (e.g. 
\pcite{honeycutt01}) and have been observed in other SW~Sex stars (e.g. BH~Lyn, 
\pcite{dhillon92}; DW~UMa, \pcite{dhillon94}; PX~And, \pcite{Still95b}). 
Low states are often accompanied by the weakening or disappearance 
of the high--excitation HeII and CIII/NIII lines, which were unfortunately 
not observed in Dec 1998. There is, however, a change in the HeI 
$\lambda$6678\AA\ Doppler maps between the two nights' observations, 
which is considered in Section~\ref{sec:trailed_spectrum}. 
Further evidence that V347~Pup exhibits changes of state is seen in the EWs 
of the emission lines between the observed epochs. For example, the EW of H$\beta$ 
varies between 17.0 $\pm$ 0.6\AA\ (July 1986, \pcite{buckley90}), 62.6 $\pm$ 1.9\AA\ 
(April 1991, \pcite{mauche94}), 9.8 $\pm$ 0.1\AA\ (Jan 1995, \pcite{still98}) and 
24.6 $\pm$ 0.2\AA\ (Jan 1998, this paper), although the high--excitation 
CIII/NIII complex has a constant EW between epochs.

We measured the phase half--width of eclipse at the out-of-eclipse 
level ($\Delta\phi$) 
by timing the first and last contacts of the eclipse and dividing by two. 
Our average value of $\Delta\phi$ = 0.110 $\pm$ 0.005 is consistent with 
the values of 0.120 $\pm$ 0.011 quoted by \scite{harrop96} and 0.105 $\pm$ 0.005 
measured by \scite{buckley90}. We then computed the radius of the accretion 
disc in V347~Pup using the geometric method outlined in \scite{dhillon91}. 
Combining $\Delta\phi$ with the system mass ratio and inclination derived in 
Section~\ref{sec:params} gives an accretion disc radius ($R_D$) of 
0.72 $\pm$ 0.09 $R_1$, where $R_1$ is the volume radius of the primary's Roche lobe. 
This value is in agreement with the value of $R_D/R_1 \ge 0.82$ quoted by 
\scite{harrop96} at the 2$\sigma$ level.

The H$\alpha$ eclipses are similar in shape to the continuum light curves, 
but do not appear to be as deeply eclipsed. The H$\beta$ and H$\gamma$ 
lines exhibit asymmetric eclipses, with ingress longer than egress. This 
behaviour is expected from asymmetric disc emission, consistent with the 
spiral arms identified in the Doppler maps (Section~\ref{sec:trailed_spectrum}). 
The high--excitation HeII + CIII/NIII complex 
has a deep and U--shaped eclipse, suggesting an origin close to the white 
dwarf. The HeI eclipses are wide with V--shaped minima, 
similar to the SW~Sex stars (e.g. \pcite{knigge00}). Note that the HeI 
flux is completely eclipsed, indicating an origin in the central 
portion of the disc, and not in an extended emission region which is larger 
than the secondary star. The HeI $\lambda$4471\AA\ emission line shows a broad 
dip in flux around phase 0.4, before climbing to reach a maximum around 
phase 0.75, which could be a further signature of the disc asymmetry.

\subsection{Trailed spectrum \& Doppler tomography}
\label{sec:trailed_spectrum}

\begin{figure*}
\begin{center}
\includegraphics[width=15cm,angle=0]{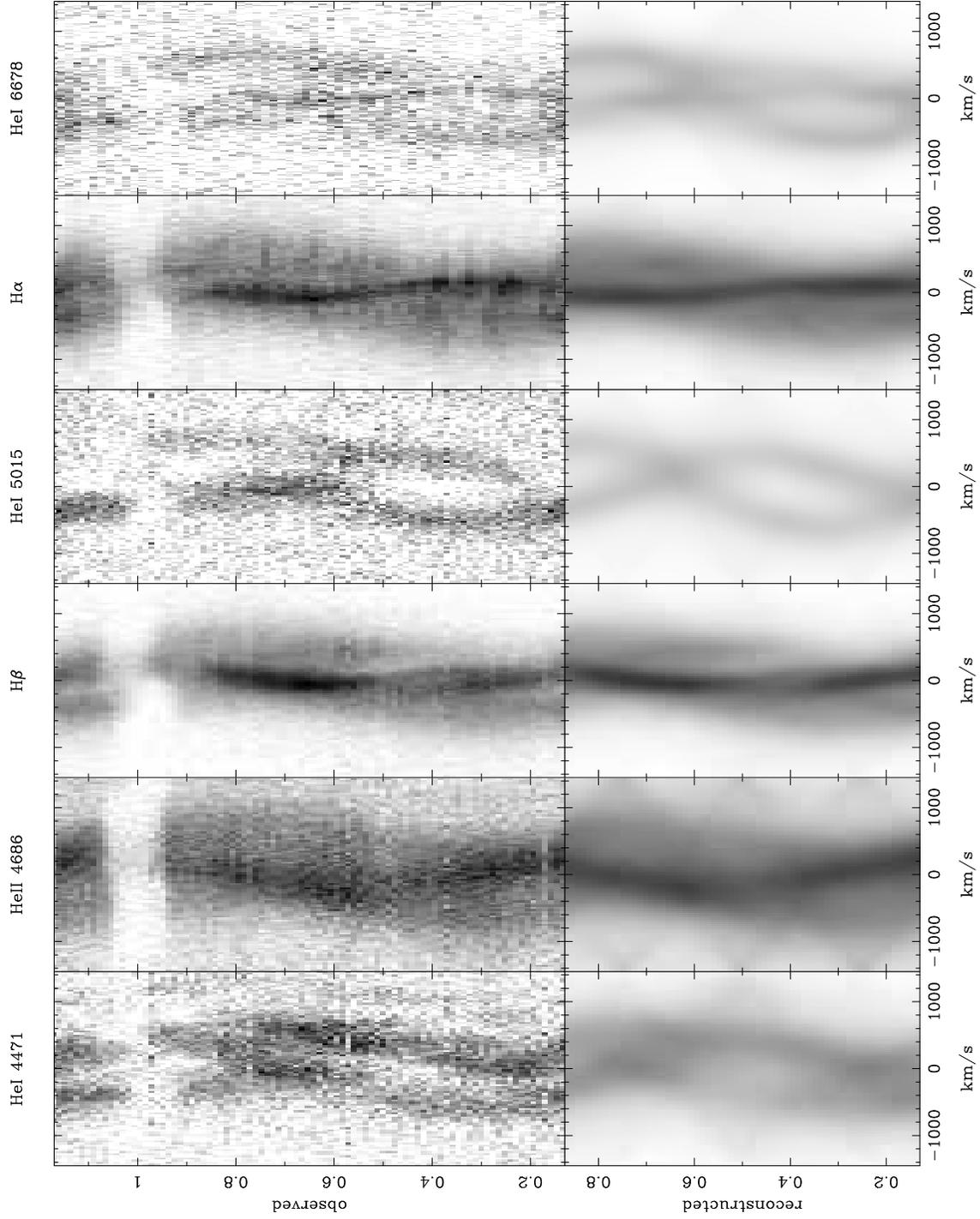}
\end{center}
\caption{
Trailed spectra and data computed from the Doppler maps 
(Fig.~\ref{fig:doppler}). The blue data 
recorded in Jan 1998 have been phase binned into 200s bins, the 
red data recorded in Dec 1998 into 300s bins. 
H$\gamma$ has not been shown, as it is very similar in nature to 
H$\beta$.}
\label{fig:trailed}
\end{figure*}

\begin{figure*}
\begin{center}
\includegraphics[width=12cm,angle=-90]{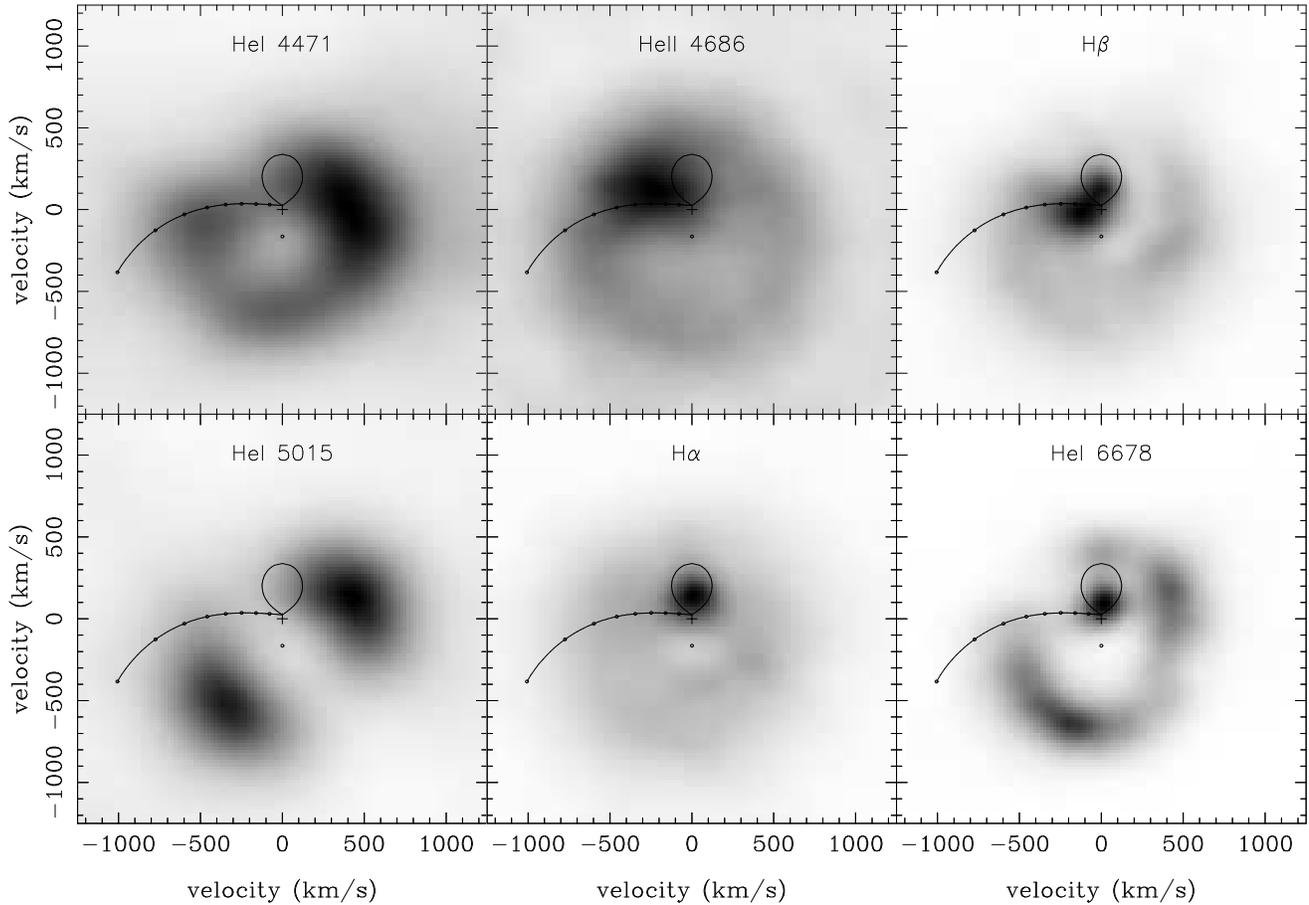}
\end{center}
\caption{
Doppler maps of the principal emission lines (H$\gamma$ is not shown, 
as it is very similar in nature to H$\beta$). The cross marked on each 
Doppler map represents the centre of mass of the system and the open circle 
represents the white dwarf. These symbols, the Roche lobe 
of the secondary star and the predicted trajectory of the gas stream, have 
been plotted using the $K_R$--corrected system parameters summarised in 
Table~\ref{tab:params}. The series of points 
along the gas stream mark the distance from the white dwarf at intervals 
of 0.1$L_1$, ranging from 1.0$L_1$ at the red star to 0.2$L_1$.
Doppler tomography cannot properly account for variable line flux, so 
spectra around primary eclipse were omitted from the fits.}
\label{fig:doppler}
\end{figure*}

We subtracted polynomial fits to the continuum and 
then rebinned the spectra onto a constant velocity--interval scale 
centred on the rest wavelength of the principal emission lines. 
For the data obtained in Jan 1998, we phase--binned all the spectra in 
order to boost the signal-to-noise (S/N). Individual spectra were weighted 
according to their S/N in order to optimally combine the spectra.
The trailed spectra of H$\alpha$, H$\beta$, HeII $\lambda$4686\AA\ and 
HeI $\lambda\lambda$4471\AA, 5015\AA, 6678\AA\ are shown in 
Fig.~\ref{fig:trailed}. 
Doppler maps were calculated for the principal emission lines 
using the modulation Doppler tomography code of \scite{steeghs03}. 
This method is an extension to the conventional Doppler tomography technique 
(e.g. \pcite{marsh00}), and maps both the constant and variable part of 
the line emission using a maximum--entropy regularised fitting procedure 
(\pcite{skilling84}). We found that the modulated contribution to the line 
emission was weak ($<1$ per cent), and thus our S/N was not sufficient 
to detect significant modulation in the accretion disc emission. We 
therefore plot in Fig.~\ref{fig:doppler} the corresponding average 
Doppler maps only. The reconstructed line profiles are plotted next to the 
observed ones in Fig.~\ref{fig:trailed} for comparison. Good fits to the data 
were achieved in all cases (reduced $\chi^2 = 1 - 1.4$)

The Balmer--line trailed spectra are dominated by a low--velocity component with a 
semi--amplitude of $\sim$150 km\,s$^{-1}$, moving from blue to red across 
primary eclipse. This is consistent with emission from the irradiated inner 
face of the secondary star, which is clearly seen in the corresponding Doppler maps. 
In the H$\beta$ map, a second low--velocity emission source is present, 
seemingly coincident with the gas stream at a distance of 0.9$L_1$, where 
$L_1$ is the distance from the white dwarf to the inner--Lagrangian point. 
There is also a weak two--armed disc asymmetry visible in the H$\beta$ emission, 
which is much more prominent in the double--peaked HeI emission lines. 
Doppler maps of V347~Pup have been produced by 
\scite{still98} for data sets recorded in 1987, 1988 and 1995. The two 
components described above from the secondary star and the disc are 
clearly visible in their maps. The summed H$\beta$ and H$\gamma$ maps 
of \scite{still98} show a stronger disc emission and spiral structure than 
our Balmer--line maps.
The disc asymmetry is significant and is reminiscent of the two armed 
spiral structures that have been observed in the discs of dwarf novae 
during outburst (e.g. \pcite{steeghs01}). We return to these in 
Section~\ref{sec:shocks}. 
The high--excitation HeII $\lambda$4686\AA\ line is dominated 
by emission from the gas stream and bright spot overlayed on a weak accretion 
disc with radius $R_D \sim 0.3 - 0.4 L_1$. Note that the HeII $\lambda$4686\AA\ 
Doppler map shows emission at higher velocities than the low--excitation 
lines, demonstrating that the material originates from closer to the white dwarf.

The blue and red HeI emission lines were recorded almost a year apart and 
exhibit clear differences in structure. The secondary star emission is 
clearly evident in HeI $\lambda$6678\AA\ (Dec 1998), although no 
strong HeI $\lambda$4471\AA\ or $\lambda$5015\AA\ emission can be 
seen in the Jan 1998 data. 
There is also a difference in the HeI $\lambda$6678\AA\ Doppler maps 
between the 25th and 27th Dec 1998, which is probably related to the change 
in brightness of the system; the 25th Dec 1998 Doppler map has more 
enhanced spiral features and weaker secondary star emission than the 
27th Dec (note that the average map of these two nights is shown in 
Fig.~\ref{fig:doppler}). 
During all these epochs, however, the spiral structures were observed, 
demonstrating that they are a persistent feature.

\subsection{Radial velocity of the white dwarf}
\label{sec:whitedwarf}

\begin{figure}
\begin{center}
\includegraphics[width=8cm,angle=0]{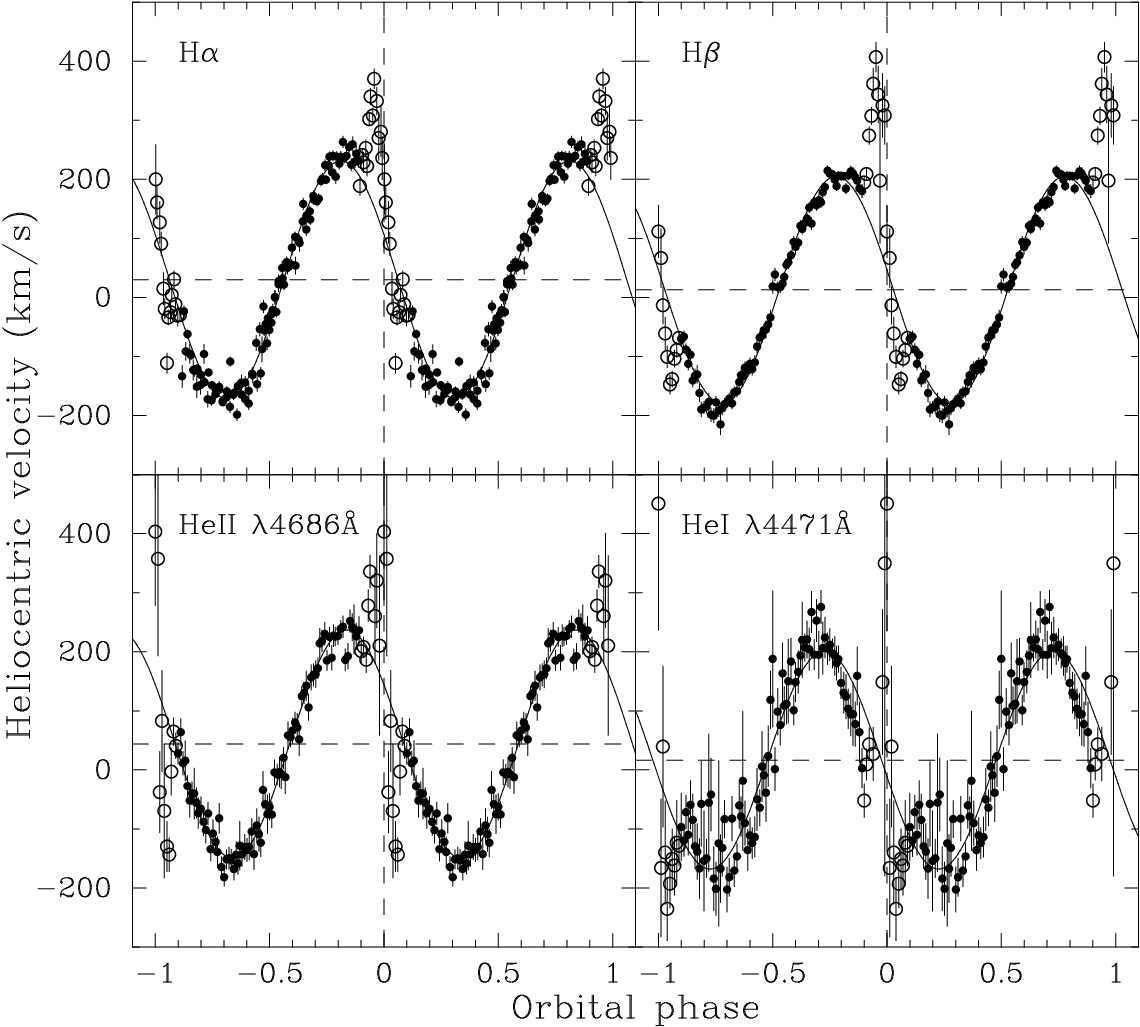} 
\end{center}
\caption{\protect\small Radial velocity curves of H$\alpha$, H$\beta$, 
HeII $\lambda$4686\AA\ and HeI $\lambda$4471\AA\ using Gaussian 
widths of 300 km\,s$^{-1}$ and a separation of 1400 km\,s$^{-1}$.
We omitted the points around primary eclipse during the 
fitting procedure (open circles) as these measurements are affected by the 
rotational disturbance. The emission lines recorded in Jan 1998 have been 
phase--binned into 100 bins for clarity.}
\label{fig:rvs}
\end{figure}

We measured the radial velocities of the emission lines in V347~Pup by applying 
the double--Gaussian method of 
\scite{schneider80}, since this technique is sensitive mainly to the 
line wings and should therefore reflect the motion of the white dwarf with 
the highest reliability. We tried Gaussians of widths 200, 300 and 
400 km\,s$^{-1}$ and we varied their separation from 200 to 3200 
km\,s$^{-1}$. We then fitted 
\begin{equation}
V=\gamma-K\sin[2\pi(\phi-\phi_0)]
\end{equation}
to each set of measurements, where $V$ is the radial velocity, $K$ the 
semi--amplitude, $\phi$ the orbital phase, and $\phi_0$ is the phase at which 
the radial velocity curve crosses from red to blue. Examples of 
the radial velocity curves measured for the H$\alpha$, H$\beta$, 
HeII $\lambda$4686\AA\ and HeI $\lambda$4471\AA\ emission lines 
are shown in Fig.~\ref{fig:rvs}. There is clear evidence of rotational 
disturbance in the emission lines, where the 
radial velocities measured just prior to eclipse are skewed to the red, and 
those measured after eclipse are skewed to the blue. This confirms the 
detection of a similar feature in the trailed 
spectra, and indicates that at least some of the emission must originate in 
the disc. There is also evidence of a phase shift in H$\alpha$ 
and HeII $\lambda$4686\AA, where the 
spectroscopic conjunction of each line occurs after photometric mid--eclipse. 
This phase shift implies an emission--line 
source trailing the accretion disc, such as a bright spot, 
and is a common feature of SW~Sex stars (e.g. DW~UMa, \pcite{shafter88}; 
V1315~Aql, \pcite{dhillon91}; SW~Sex, \pcite{dhillon97b}). There appear 
to be no significant phase shifts, however, in the other emission lines. 
\scite{buckley90}, \scite{mauche94} and 
\scite{diaz99} find no evidence of phase shift in any of their emission lines, 
although their errors on $\phi_0$ were much larger.

\begin{figure}
\begin{center}
\includegraphics[width=5.4cm,angle=-90]{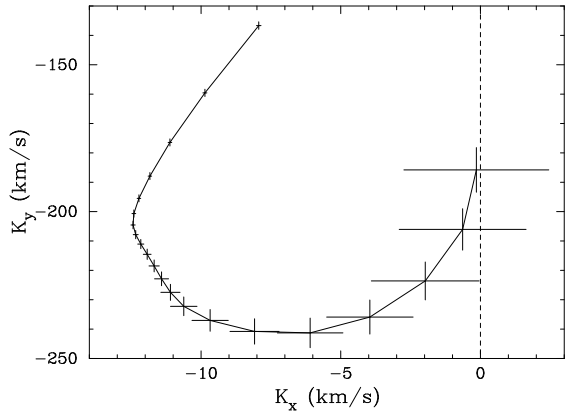}\\
\end{center}
\caption{\protect\small
Light--centres diagram for H$\alpha$. Points are plotted for radial velocity 
fits using Gaussians of FWHM = 300 km\,s$^{-1}$, with separations from 
900 km\,s$^{-1}$ to 2900 km\,s$^{-1}$ at 100 km\,s$^{-1}$ intervals. The points 
move anti--clockwise, towards the $K_x = 0$ axis with increasing Gaussian 
separation.}
\label{fig:centres}
\end{figure}

We tried to measure white dwarf radial velocity ($K_W$) values using a 
diagnostic diagram (\pcite{shafter86}), but with no success. 
We therefore attempted to make use of the light--centres method, 
as described by \scite{marsh88a}. In the co-rotating co-ordinate system, 
the white dwarf has velocity ($0, -K_W$), and symmetric emission, say 
from a disc, would be centred at that point. By plotting 
$K_x = -K\sin\phi_0$ versus $K_y = -K\cos\phi_0$ for the different 
radial velocity fits (Fig.~\ref{fig:centres}), one finds that the points move 
closer to the $K_y$ axis with increasing Gaussian separation. A simple 
distortion which only affects low velocities, such as a bright spot, would 
result in this pattern, equivalent to a decrease in distortion as one measures 
emission further into the line wings and therefore more closely representing 
the velocity of the primary star. By linearly extrapolating the largest Gaussian 
separation on the H$\alpha$ light--centre diagram to the 
$K_y$ axis, we measure the radial velocity semi--amplitude of the white 
dwarf to be $\sim$180 km\,s$^{-1}$. The large uncertainty in this value 
($\sim$ 40 km\,s$^{-1}$), however, and the unsuccessful application of the 
technique to the other emission lines, prompted us to proceed with the mass 
determination using the secondary star features alone.

%\begin{figure}
%\begin{center}
%\begin{tabular*}{85mm}{l}
%
%(a)\\
%\\
%\includegraphics[width=8.5cm,angle=0]{v347pup_diag300.ps}\\
%\\
%(b)\\
%\includegraphics[width=6cm,angle=-90]{v347pup_ha_centres_paper.ps}\\
%\end{tabular*}
%\end{center}
%\caption{\protect\small
%(a) Diagnostic diagram for V347~Pup based on the double--Gaussian 
%radial velocity fits to H$\alpha$ and HeI $\lambda$6678\AA. Each Gaussian 
%has a FWHM of 300 km\,s$^{-1}$. 
%(b) Light--centres diagram for H$\alpha$.}
%\label{fig:diag}
%\end{figure}

\subsection{Radial velocity of the secondary star}
\label{sec:secondary}

The secondary star in V347~Pup is clearly visible in 
Fig.~\ref{fig:av_spec} through absorption lines of 
MgI, FeI and CaI. 
We compared regions of the spectra rich in absorption lines with 
a number of templates with spectral types G7V--M3.5V. A technique 
known as skew mapping was used to enhance the secondary features and obtain a 
measurement of the radial velocity semi--amplitude of the secondary star ($K_R$). 
See \scite{vandeputte03} for a detailed 
critique of skew mapping and \scite{thoroughgood04} for a successful 
application to AC~Cnc and V363~Aur. 

The data centred on $\lambda$5290\AA\ were recorded specifically to 
exploit the secondary star 
features found between the H$\beta$ and H$\alpha$ lines. Unfortunately, 
the presence of weak emission lines (e.g. FeII multiplet 42 at 
$\lambda\lambda$4924\AA, 5018\AA\ and 5169\AA, \pcite{mason03}) 
hampered all efforts to determine a $K_R$ value from these data. 
The dominance of the emission lines in the spectra 
centred on $\lambda$4610\AA\ also prevented a $K_R$ determination from 
these data. The red spectra of V347~Pup centred on $\lambda$6330\AA, 
however, allowed us to 
study the secondary star through absorption features blueward of H$\alpha$, 
such as the CaI $\lambda$6162\AA$\:$ line. Exactly the same conclusion 
was reached by \scite{diaz99}.

The first step was to shift the spectral type template stars to correct for 
their radial velocities. We then normalized each spectrum by dividing 
by a constant and then subtracting a polynomial fit to the continuum. 
This ensures that line strength is preserved along the 
spectrum. The V347~Pup spectra were normalized in the same way. 
The template spectra were artificially broadened to account for both the 
orbital smearing of the V347~Pup spectra due to their exposure times ($t_{exp}$), 
using the formula
\begin{equation}
{V} =  {{{t_{exp}}{2\pi}{K_R}} \over {P}}
\label{eqn:smear}
\end{equation}
(e.g. \pcite{watson00}), and the rotational velocity of the 
secondary ($v \sin i$). Estimated values of $K_R$ and $v \sin i$ were used 
in the first instance, before iterating to find the best--fitting values given 
in Section~\ref{sec:params}.
Regions of the spectrum devoid of emission lines were then cross--correlated 
with each of the templates, yielding a time series of cross--correlation 
functions (CCFs) for each template star. The regions used for the 
cross--correlation can be seen in Fig.~\ref{fig:residual}.
To produce the skew maps, these CCFs were then back--projected in the 
same way as time--resolved spectra in standard 
Doppler tomography (\pcite{marsh88b}). If there is a detectable 
secondary star, we expect a peak at (0,$K_R$) in the skew map.
This can be repeated for each of the templates, and the final skew map is the 
one that gives the strongest peak.

\begin{figure}
\begin{center}
\includegraphics[width=7.5cm,angle=0]{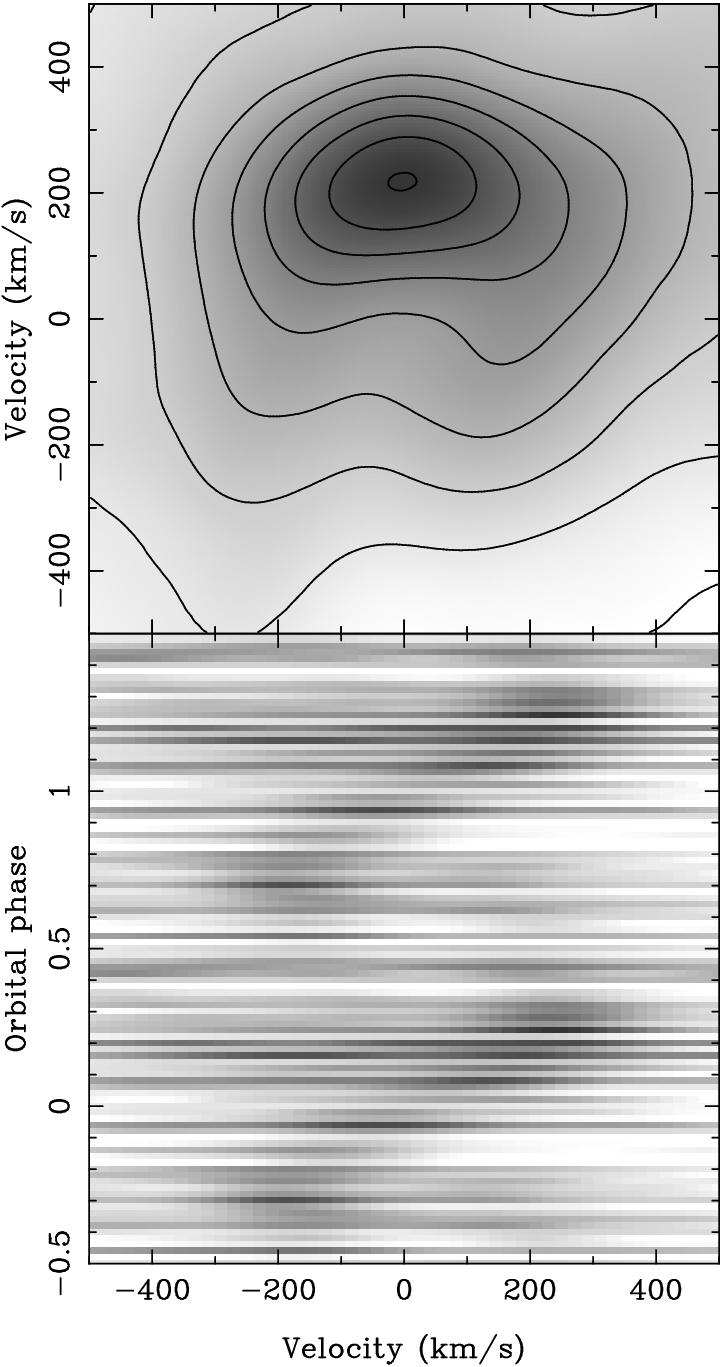}
\end{center}
\caption{\protect\small
Skew maps (top) and trailed CCFs (bottom) of V347~Pup cross--correlated with 
a M0.5V dwarf template.}
\label{fig:skewmaps}
\end{figure}

The skew maps show well--defined peaks at 
$K_y \approx$ 216 km\,s$^{-1}$ -- the skew map of the M0.5V template is 
shown in Fig.~\ref{fig:skewmaps} together with the trailed CCFs. 
A systemic velocity of $\gamma$ = 15 km\,s$^{-1}$ was applied in order to 
shift the skew map peaks onto the $K_x$ = 0 axis (see \pcite{sad98b} for details). 
We therefore adopt $\gamma$ = 15 $\pm$ 5 km\,s$^{-1}$ as the 
systemic velocity of V347~Pup, in excellent agreement with the values of 
16 $\pm$ 10 km\,s$^{-1}$ and 15 $\pm$ 12 km\,s$^{-1}$ measured by 
\scite{still98} from the Balmer and HeII $\lambda$4686\AA$\:$ emission lines. 
The $\gamma$ velocities from the emission lines shown in Fig.~\ref{fig:rvs} 
ranged between 13 km\,s$^{-1}$ and 44 km\,s$^{-1}$.
Other $\gamma$ values measured from optical emission lines 
vary widely in the literature (--3 to 60 km\,s$^{-1}$, \pcite{diaz99}; --9 to 
159 km\,s$^{-1}$, \pcite{mauche94}). 

Our adopted $K_R$ of 216 $\pm$ 5 km\,s$^{-1}$ was derived from the skew map 
peak of the best--fitting template found in Section~\ref{sec:params}. This result 
actually covers the 
$K_R$ values derived from $all$ of the template stars to within 
the errors, demonstrating that the result is robust to the choice of 
template (see Table~\ref{tab:vsini}). 

\subsection{Rotational velocity and spectral type of the secondary star}
\label{sec:rotational}

The spectral--type templates were broadened for smearing due to orbital 
motion, as before, and rotationally broadened by a range of velocities 
(50--240 km\,s$^{-1}$). We then ran an optimal subtraction routine, which 
subtracts a constant times the normalized template spectrum from the 
normalized average V347~Pup spectrum, adjusting the constant to 
minimize the scatter in the residual. (Normalisation was carried out in the 
same way as Section~\ref{sec:secondary}, except that this time, the spectra 
were set to unity.) The scatter is measured 
by carrying out the subtraction and then computing the $\chi^2$ between the 
residual spectrum and a smoothed version of itself. By finding the value of 
rotational broadening that minimizes the $\chi^2$, we obtain an 
estimate of both $v \sin i$ and the spectral type of the secondary star. 
Note that the $v \sin i$ values of the template stars are much lower than 
the instrumental resolution, so do not affect our measurements of 
$v \sin i$ for the secondary star.

The value of $v \sin i$ obtained using this method varies depending on the 
spectral type template, the wavelength region for optimal subtraction, 
the amount of smoothing of the residual spectrum in the calculation of 
$\chi^2$ and the value of the limb--darkening coefficient used in the 
broadening procedure. 
The values of $v \sin i$ for all of the templates calculated using 
values for the limb--darkening coefficient of 0.5 and smoothed using 
a Gaussian of FWHM = 15km\,s$^{-1}$, are listed in Table~\ref{tab:vsini}. 

\begin{figure}
\begin{center}
\includegraphics[width=6.2cm,angle=-90]{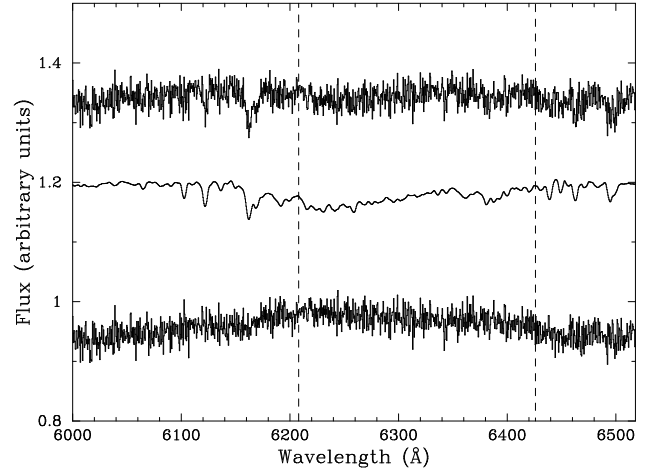}
\end{center}
\caption{\protect\small
Orbitally--corrected average spectrum of V347~Pup (top) with the 
broadened M0.5V template (middle) and the residuals after optimal 
subtraction (bottom). The template spectrum has been multiplied by 
the scaling factor found from the optimal subtraction. All of the spectra 
are normalised and offset on the plot by 
an arbitrary amount for clarity. The wavelength limits shown are those used 
for the cross--correlation and optimal subtraction procedures, except for the 
region between the dashed lines owing to few secondary star features.}
\label{fig:residual}
\end{figure}

\begin{figure}
\begin{center}
\begin{tabular}{c}
\includegraphics[width=7.5cm,angle=0]{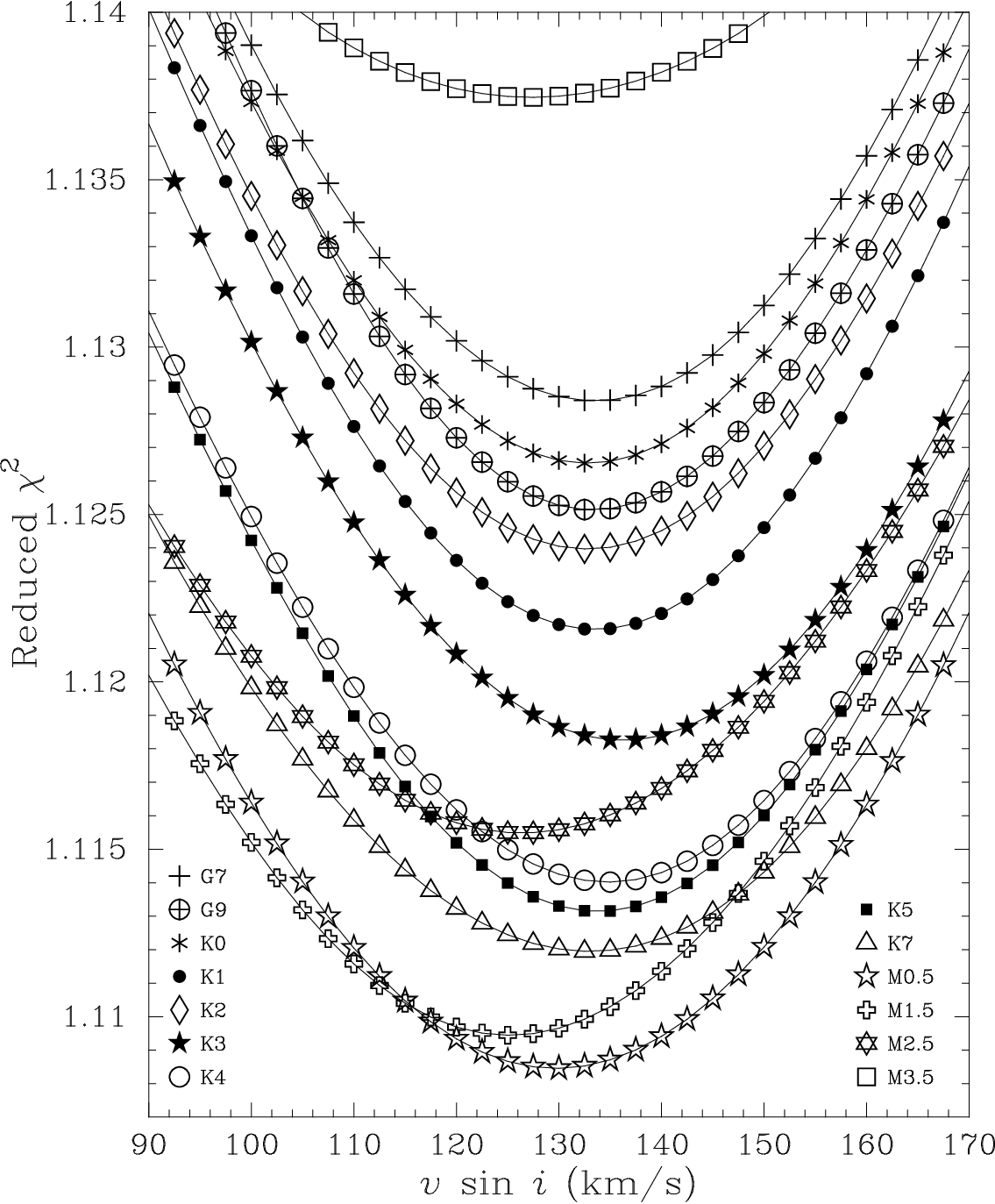}
\end{tabular}
\end{center}
\caption{Determination of $v \sin i$ for V347~Pup 
using different spectral--type templates. Degrees of freedom = 699.}
\label{fig:vsini}
\end{figure}

A plot of $\chi^2$ versus $v \sin i$ for each spectral--type template is 
shown in Fig.~\ref{fig:vsini}. The spectral type with the lowest 
$\chi^2$ value is M0.5V, which agrees with a visual identification 
of the best fitting template. \scite{diaz99}, however, 
estimate a secondary star spectral type between K0V and K5V, with the 
possibility of a later--type subgiant.
A plot of the V347~Pup average spectrum, a broadened M0.5V template spectrum and 
the residual of the 
optimal subtraction is shown in Fig.~\ref{fig:residual}. The $\chi^2$ 
for the M0.5V template has a minimum at 130 km\,s$^{-1}$, so 
we adopt $v \sin i$ = 130 $\pm$ 5 km\,s$^{-1}$, with the error accounting for 
the measurement accuracy and the other variables noted in the 
previous paragraph. The error quoted on our adopted value encompasses the 
measured $v \sin i$ 
for all of the templates used in the analysis (except for K3V with 
$v \sin i$ = 136 km\,s$^{-1}$).

\begin{table}
{\protect\small
\caption{$v \sin i$ values for V347~Pup cross--correlated with 
the rotationally broadened profiles of G7 -- M3.5V templates.
Also shown is the factor used to multiply the template star features during 
optimal extraction, and the position of the strongest peak in the 
skewmaps derived from each template using $\gamma$--velocities 
of 0 km\,s$^{-1}$ and 15 km\,s$^{-1}$.}
\label{tab:vsini}
\begin{tabular}{lcr@{$\:\pm\:$}lr@{,}lr@{,}l}
\hline
\vspace{-3mm}\\
\multicolumn{1}{c}{Templates} & 
\multicolumn{1}{c}{$v \sin i$} & 
\multicolumn{2}{c}{Optimal} & 
\multicolumn{2}{c}{($K_x,K_y$)} & 
\multicolumn{2}{c}{($K_x,K_y$)} \\
\multicolumn{1}{c}{} & 
\multicolumn{1}{c}{at min $\chi^{2}$} & 
\multicolumn{2}{c}{factor} &
\multicolumn{2}{c}{$\gamma=0$} &
\multicolumn{2}{c}{$\gamma=15$} \\
\multicolumn{1}{c}{} & 
\multicolumn{1}{c}{(km\,s$^{-1}$)} & 
\multicolumn{2}{c}{} &
\multicolumn{2}{c}{(km\,s$^{-1}$)} &
\multicolumn{2}{c}{(km\,s$^{-1}$)} \\
\vspace{-3mm}\\
\hline
\vspace{-3mm}\\
G7V & 134 & 0.32&0.05 & (--26&212) & (10&220) \\
G9V & 133 & 0.28&0.04 & (--13&215) & (13&220) \\
K0V & 133 & 0.23&0.03 & (--2&217) & (14&219) \\
K1V & 134 & 0.24&0.03 & (--15&215) & (8&220) \\
K2V & 133 & 0.20&0.03 & (--22&212) & (6&219) \\
K3V & 136 & 0.19&0.03 & (--28&212) & (0&217) \\
K4V & 135 & 0.14&0.02 & (--17&211) & (3&217) \\
K5V & 134 & 0.13&0.02 & (--17&213) & (1&218) \\
K7V & 133 & 0.12&0.02 & (--24&210) & (--3&216) \\
M0.5V & 130 & 0.13&0.02 & (--18&213) & (0&216) \\
M1.5V & 125 & 0.12&0.02 & (--17&213) & (--2&216) \\
M2.5V & 126 & 0.13&0.02 & (--21&213) & (--7&216) \\
M3.5V & 127 & 0.12&0.02 & (--33&213) & (--23&217) \\
\hline\\
\end{tabular}
}
\end{table}

%\begin{table}
%{\protect\small
%\caption{$v \sin i$ values for V347~Pup cross--correlated with 
%the rotationally broadened profiles of G7 -- M3.5V templates.
%The third and fourth columns list the position of the strongest peak in the 
%skewmaps derived from each template using $\gamma$--velocities 
%of 0 km\,s$^{-1}$ and 15 km\,s$^{-1}$.}
%\label{tab:vsini}
%\begin{center}
%\begin{tabular}{lcr@{,}lr@{,}l}
%\hline
%\vspace{-3mm}\\
%\multicolumn{1}{c}{Templates} & 
%\multicolumn{1}{c}{$v \sin i$} & 
%\multicolumn{2}{c}{($K_x,K_y$) from} &
%\multicolumn{2}{c}{($K_x,K_y$) from} \\
%\multicolumn{1}{c}{} & 
%\multicolumn{1}{c}{at min $\chi^{2}$} & 
%\multicolumn{2}{c}{$\gamma=0$ skew map} &
%\multicolumn{2}{c}{$\gamma=15$ skew map} \\
%\multicolumn{1}{c}{} & 
%\multicolumn{1}{c}{(km\,s$^{-1}$)} & 
%\multicolumn{2}{c}{(km\,s$^{-1}$)} & 
%\multicolumn{2}{c}{(km\,s$^{-1}$)} \\
%\vspace{-3mm}\\
%\hline
%\vspace{-3mm}\\
%G7V & 134 & (--26&212) & (10&220) \\
%G9V & 133 & (--13&215) & (13&220) \\
%K0V & 133 & (--2&217) & (14&219) \\
%K1V & 134 & (--15&215) & (8&220) \\
%K2V & 133 & (--22&212) & (6&219) \\
%K3V & 136 & (--28&212) & (0&217) \\
%K4V & 135 & (--17&211) & (3&217) \\
%K5V & 134 & (--17&213) & (1&218) \\
%K7V & 133 & (--24&210) & (--3&216) \\
%M0.5V & 130 & (--18&213) & (0&216) \\
%M1.5V & 125 & (--17&213) & (--2&216) \\
%M2.5V & 126 & (--21&213) & (--7&216) \\
%M3.5V & 127 & (--33&213) & (--23&217) \\
%\hline\\
%\end{tabular}
%\end{center}
%}
%\end{table}

\subsection{The $K_R$ correction}
\label{sec:kcor}

The irradiation of the secondary stars in CVs by the emission regions around 
the white dwarf and the bright spot has been shown to influence the measured 
$K_R$ (e.g. \pcite{wade88}, \pcite{watson00}). For example, if absorption 
lines are quenched on the irradiated side of the secondary, the 
centre of light will be shifted towards the back of the star. The measured 
$K_R$ will then be larger than the true (dynamical) value. 

\scite{diaz99} found evidence for irradiation of the secondary star 
in V347~Pup, leading them to apply a correction to their measured 
$K_R$ value. This fact, and the presence of Balmer and HeI emission 
from the inner face of the secondary star seen in the Doppler maps and 
trailed spectra (Section~\ref{sec:trailed_spectrum}), prompted us to look 
for similar irradiation effects 
in the absorption lines of our data. We applied the following two observational 
tests. First, the rotationally broadened line profile would be distorted if 
there was a non--uniform absorption distribution across the surface of the 
secondary star (\pcite{davey92}). This would result in a non--sinusoidal 
radial velocity curve. Second, one would expect a depletion of secondary star 
absorption--line 
flux at phase 0.5, where the quenched inner--hemisphere is pointed towards 
the observer (e.g. \pcite{friend90a}). 

\begin{figure}
\begin{center}
\begin{tabular}{c}
\includegraphics[width=8cm,angle=0]{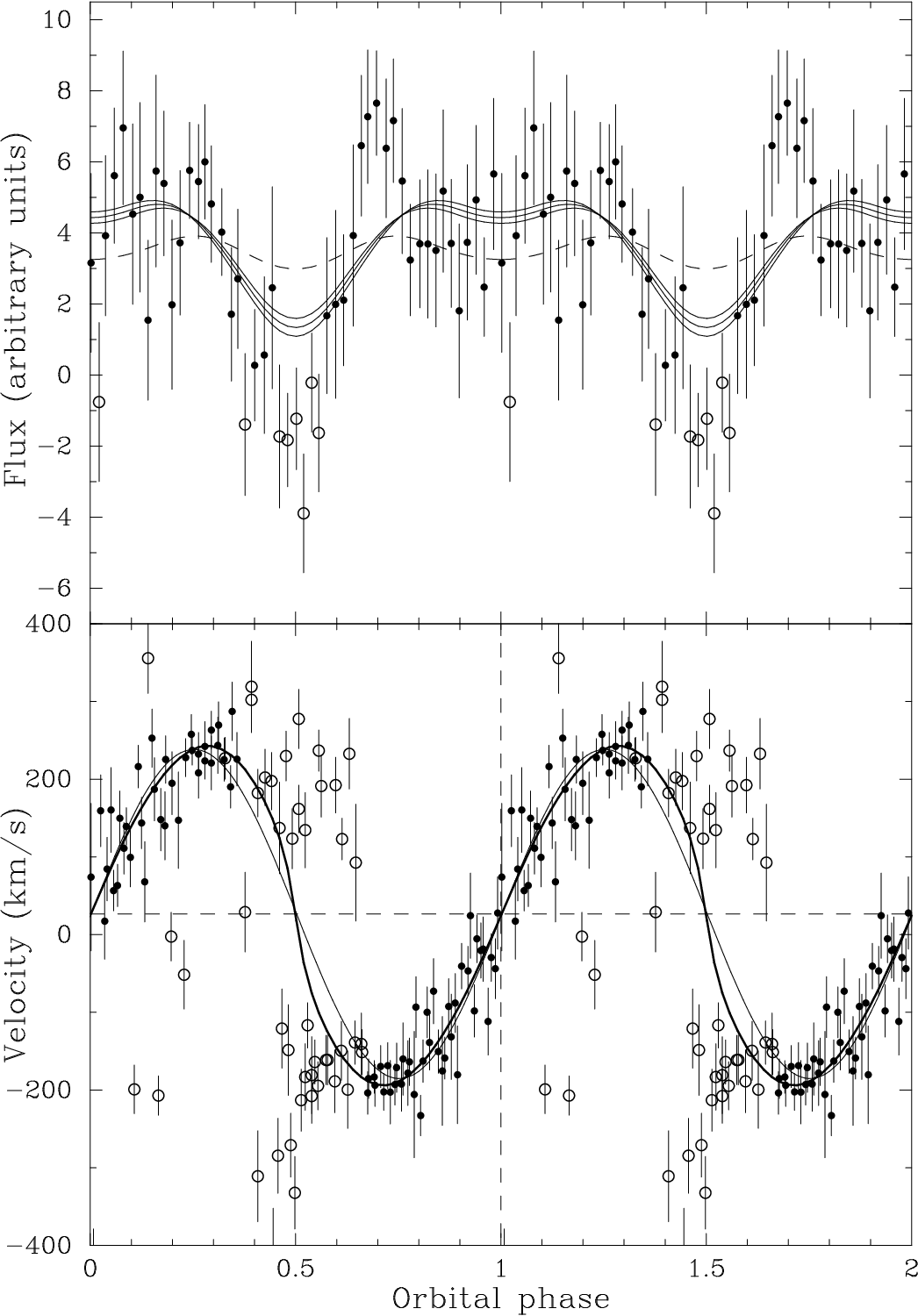}\\
\end{tabular}
\end{center}
\caption{
Upper panel: Secondary star absorption line light curve with model fits 
(solid lines). Model fits are shown for Roche lobes with 11, 12 
and 13 slices removed (see text for details). The lower the 
line, the more slices removed. The dashed line represents a model 
where 0 slices have been removed. The data have been phase--binned into 
50 bins to increase S/N. 
Lower panel: Measured secondary star radial velocity curve with a sinusoidal fit 
(thin solid line) and the best--fitting model fit (thick solid line). 
In both panels, the open circles indicate points that have been omitted from 
the fits and the data have been folded to show 2 orbital phases.}
\label{fig:irrplot}
\end{figure}

The secondary star radial velocity curves were produced by cross--correlating 
the V347~Pup spectra with the best--fitting smeared and broadened template 
spectra, as described in Section~\ref{sec:secondary}. The 
cross--correlation peaks were plotted against phase to produce the radial 
velocity curves shown in the lower panel of Fig.~\ref{fig:irrplot}. 
There is evidence for an eccentricity in the radial velocity curve compared 
with the sinusoidal fit represented by the thin solid line, 
although the data are noisy. 

The variation of secondary star absorption--line flux with phase for 
V347~Pup is shown in the top panel of Fig.~\ref{fig:irrplot}. 
These light curves were produced by 
optimally subtracting the smeared and rotationally broadened 
best--fitting template from the individual CV spectra (with the secondary 
radial velocity shifted out) as described in Section~\ref{sec:rotational}. 
This time, however, the 
spectra were continuum subtracted rather than normalised to ensure that the 
measurements were not affected by a fluctuating disc brightness. The constants 
produced by the optimal subtraction are secondary star absorption--line fluxes, 
correct relative 
to each other, but not in an absolute sense. 
The dashed lines super--imposed on the light curves represent the variation of 
flux with phase for a Roche lobe with a uniform absorption distribution. The 
sinusoidal nature is the result of the changing projected area of the 
Roche lobe through the orbit. The V347~Pup light curve is clearly not represented 
by a uniform Roche lobe distribution as the secondary star absorption--line 
flux vanishes between phases 0.4--0.6.

These three pieces of evidence, as well as the disappearance of the CCFs 
between phases 0.4--0.6 seen in Fig.~\ref{fig:skewmaps} suggest that the 
secondary star in V347~Pup is 
irradiated and we must correct the $K_R$ values accordingly.

It is possible to correct $K_R$ for the effects of irradiation by 
modelling the secondary star flux distribution. In our simple model, we divided 
the secondary Roche lobe into 40 vertical slices of equal width from the 
$L_1$ point to the back of the star. 
We then produced a series of model light curves (using the system parameters 
derived in Section~\ref{sec:params}), varying the numbers of 
slices omitted from the inner hemisphere of the secondary 
which contribute to the total flux. 
The model light curves were then scaled to match the observed data, and the 
best--fitting model found by measuring the $\chi^2$ between the two.
In all models, we used a gravity--darkening parameter $\beta = 0.08$ and 
limb--darkening coefficient $u = 0.5$ (e.g. \pcite{watson00}). 
The negative data points around phase 0.5 were set to zero, as the 
secondary star absorption line flux disappears at this point. 
Once the best--fitting light curve was found, 
we produced fake V347~Pup spectra from the model, which were 
cross-correlated with a fake template star to produce a synthetic radial 
velocity curve. 
In the first instance, the synthetic curve mimicked the non-sinusoidal 
nature of the observed data, but with a larger semi--amplitude. This was 
expected, as the model input parameters used the uncorrected 
$K_R$ derived in Section~\ref{sec:params}. 
We then lowered $K_R$ and repeated 
the process, until the 
semi--amplitude of the model and observed radial velocity curves were in 
agreement, each time checking the light curve models for goodness of fit. 
The resulting $K_R$ was 
then adopted as the real (or dynamical) $K_R$ value. 

The best--fitting model light curve was produced by omitting 
12 slices when fitting the data (reduced $\chi^2$ between model and data = 
1.03). The model light curves omitting 
11, 12 and 13 slices are shown by the solid lines in Fig.~\ref{fig:irrplot}. 
Our final model, which has an input $K_R$ of 198 km\,s$^{-1}$, produces the 
radial velocity curve shown as the thick solid line in the lower 
panel of Fig.~\ref{fig:irrplot}.
There is good agreement between this and the observed data. 
If gravity--darkening and limb--darkening are 
neglected, the best fit light curve remains the same, but produce a $K_R$ 
value which is $\sim$ 6 km\,s$^{-1}$ lower.

In summary, we correct the $K_R$ of V347~Pup from 216 km\,s$^{-1}$ 
to 198 km\,s$^{-1}$. This correction of 18 km\,s$^{-1}$ is exactly the same 
as that calculated by \scite{diaz99} using a much simpler approximation, 
which changed their measured value of 205 km\,s$^{-1}$ to 187 km\,s$^{-1}$.

\subsection{The distance to V347~Pup}
\label{sec:distance}

By finding the apparent magnitude of the secondary star from its 
contribution to the total light of the system, and estimating its 
absolute magnitude, we can calculate the distance ($d$) using the equation: 

\begin{equation}
\label{eqn:distance}
{5 \log(d/10) = m_V - M_V - d\;A_V/1000}
\end{equation}

\noindent{where $A_V$ is the visual interstellar extinction in 
magnitudes per kpc.}

The mean $R$--band photometric flux of V347~Pup during the recorded spectra 
is 8.93 mJy, which we convert to a mean $R$--band 
magnitude of 13.8 $\pm$ 0.3. The uncertainty reflects the 
change in brightness of the system between 25 and 27 Dec. During this time, 
the secondary star contributes 13 $\pm$ 2 per cent of the total light of the 
system, assuming an early M spectral type (see Table~\ref{tab:vsini}). 
The apparent magnitude of the secondary is therefore $R$ = 
16.0 $\pm$ 0.4, which we convert to a $V$--band magnitude of 17.3 $\pm$ 0.4 
using a typical $V-R$ value for an early M star from \scite{gray92}.
There are a number of ways of estimating the absolute magnitude of the 
secondary star, assuming it is on the main sequence (e.g. 
\pcite{patterson84}; \pcite{warner95b}; \pcite{gray92}). We took each 
of these into account and adopt an average value of $M_V$ = +8.8 $\pm$ 
0.5. \scite{mauche94} estimated the extinction to V347~Pup to be 
E($B-V$) = 0.05, which results in $A_V = 0.16$ (\pcite{scheffler82}). The 
distance to V347~Pup is calculated from equation~\ref{eqn:distance} 
to be 490 $\pm$ 130 pc.

\scite{buckley90} estimate the distance to V347~Pup to be between 174--380 
pc, based on their measured system inclination and out-of-eclipse magnitude. 
\scite{mauche94} use their interstellar reddening measurement and a 
mean interstellar hydrogen number density to estimate a distance of 340--590 pc. 
Finally, \scite{diaz99} find a distance of 510 $\pm$ 160 pc from the spectral line 
depths of the secondary star. Our value is consistent with all 
distance estimates in the literature.

\subsection{System parameters}
\label{sec:params}

Using the $K_R$ and $v \sin i$ values found in Sections~\ref{sec:rotational} 
and~\ref{sec:kcor} in conjunction with the period determined in 
Section~\ref{sec:ephem} and a measurement of the 
eclipse full--width at half depth ($\Delta\phi_{1/2}$), we can calculate 
accurate system parameters for V347~Pup.

In order to determine $\Delta\phi_{1/2}$, we estimated the flux out of 
eclipse (the principal source of error) and at eclipse minimum, 
and then measured the full--width of the eclipse half-way between these 
points. The eclipse full--width at half-depth was measured to 
be $\Delta\phi_{1/2}$ = 0.115 $\pm$ 0.005, in agreement with the eclipse 
half--width at half--depth of 0.052 $\pm$ 0.002 measured by \scite{buckley90} 
at the 2$\sigma$ level.

We have opted for a Monte Carlo approach similar to \scite{horne93} to 
calculate the system parameters and their errors. For a given set 
of $K_R$, $v \sin i$, $\Delta\phi_{1/2}$ and $P$, the other system parameters 
are calculated as follows.

$R_2/a$ can be estimated because we know that the secondary star fills its 
Roche lobe (as there is an accretion disc present and hence mass transfer). 
$R_2$ is the equatorial radius of the secondary star and $a$ is the binary 
separation. We used Eggleton's formula (\pcite{eggleton83}) which gives the 
volume-equivalent radius of the Roche lobe to better than 1 per cent, which 
is close to the equatorial radius of the secondary star as seen during 
eclipse,
\begin{equation}
{{R_2} \over a} = {{0.49q^{2/3}} \over {{0.6q^{2/3} + \ln{(1+q^{1/3})}}}}. 
\label{eqn:eggleton}
\end{equation}
The secondary star rotates synchronously with the orbital motion, so 
we can combine $K_R$ and $v \sin i$, to get
\begin{equation}
{{R_2} \over a}{(1 + q)} = {{v \sin i} \over {K_R}}.
\label{eqn:synch}
\end{equation}
By considering the geometry of a point eclipse by a spherical body 
(e.g. \pcite{dhillon91}), the radius of the secondary can be shown to be
\begin{equation}
\biggl({{R_2} \over a}\biggr)^2 = \sin^2\pi\Delta\phi_{1/2}+
\cos^2\pi\Delta\phi_{1/2}\cos^2i,
\label{eqn:inclin}
\end{equation}
which, using the value of $R_2/a$ obtained using equations~\ref{eqn:eggleton} 
and~\ref{eqn:synch}, allows us to calculate the inclination, $i$, of the system. 
The geometry of a disc eclipse can be approximated to a point eclipse if the 
light distribution around the white dwarf is axi--symmetric (e.g. \pcite{dhillon90}). 
This approximation is justified given the symmetry of the primary eclipses 
in the photometry light curves (Fig.~\ref{fig:lc}). 
Kepler's Third Law gives us
\begin{equation}
{{K_R^3P_{orb}}\over{2\pi G}}={{M_1\sin^3i}\over{(1+q)}^2},
\end{equation}
which, with the values of $q$ and $i$ calculated using equations 
~\ref{eqn:eggleton},~\ref{eqn:synch} and~\ref{eqn:inclin}, gives the mass 
of the primary star. The mass of the secondary star can then be obtained using 
\begin{equation}
{M_2} = {q{M_1}}.
\label{eqn:qratio}
\end{equation}
The radius of the secondary star is obtained from the equation 
\begin{equation}
{{v \sin i} \over {R_2}} = {{2\pi \sin i} \over P},
\label{eqn:secradius}
\end{equation}
(e.g. \pcite{warner95a}) and the separation of the components, $a$, 
is calculated from equations 
~\ref{eqn:synch} and ~\ref{eqn:secradius} with $q$ and $i$ now known.

\begin{figure}
\begin{center}
\includegraphics[width=7.8cm,angle=-90]{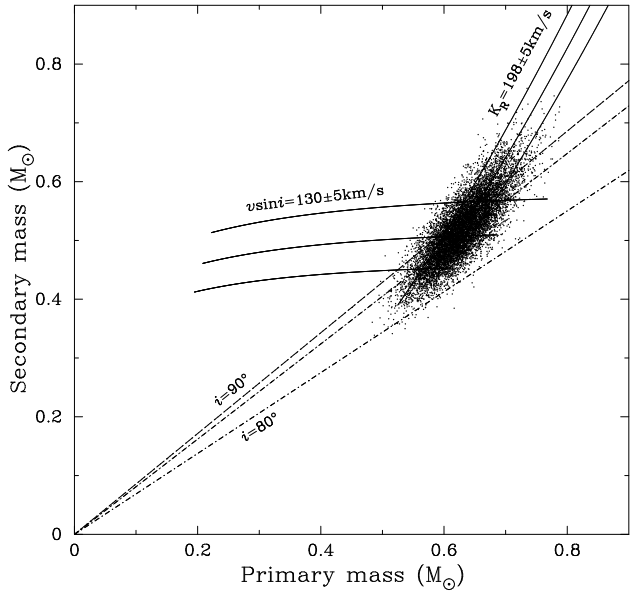}
\end{center}
\caption{\protect\small
Monte Carlo determination of system parameters for V347~Pup. 
Each dot represents 
an $M_1,M_2$ pair; the solid 
curves satisfy the $v \sin i$ and $K_R$ constraints, and the dashed lines 
mark lines of constant inclinations ($i$ = 80$^\circ$, 85$^\circ$ and 
90$^\circ$).}
\label{fig:montecarlo}
\end{figure}

The Monte Carlo simulation takes 10\,000 values of $K_R$, $v \sin i$, and 
$\Delta\phi_{1/2}$ (the error on the period is deemed to be negligible in 
comparison to the errors on $K_R$, $v \sin i$, and $\Delta\phi_{1/2}$), 
treating each as being normally distributed about their 
measured values with standard deviations equal to the errors on the 
measurements. We then calculate the masses of the components, the inclination 
of the system, the radius of the secondary star, and the separation of the 
components, as outlined above, omitting 
($K_R$, $v \sin i$, $\Delta\phi_{1/2}$) 
triplets which are inconsistent with $\sin i \leq1$. Each accepted 
$M_1,M_2$ pair is then plotted as a point in Figure~\ref{fig:montecarlo}, 
and the masses and their errors are computed from 
the mean and standard deviation of the distribution of these pairs. 

We find the component masses of V347~Pup to be $M_1 = 0.63 \pm 0.04M_\odot$ 
and $M_2 = 0.52 \pm 0.06M_\odot$.
The values of all the system parameters deduced from the Monte Carlo 
computation are listed in Table~\ref{tab:params}, including $K_R$--corrected 
and non $K_R$--corrected values for comparison. Note that our derived 
$K_W$ of 163 $\pm$ 9 km\,s$^{-1}$ is in 
remarkable agreement with the $K_W$ values of \scite{still98} who measure 
156 $\pm$ 10 km\,s$^{-1}$ using a double--Gaussian convolution of the 
Balmer lines, and 166 km\,s$^{-1}$ as the centre of axisymmetric Balmer emission. 
The white dwarf mass of 0.63 $\pm$ 0.04$M_\odot$ is consistent 
with the average value of ${\overline M}_1 = 0.80\pm0.22M_\odot$ (for 
CVs above the period gap) determined by \scite{smith98}.
The empirical relation obtained by \scite{smith98} between mass and radius 
for the secondary stars in CVs predicts that if the secondary star in 
V347~Pup is on the main-sequence, it should have a radius of 
0.54 $\pm$ 0.08 $R_\odot$. Our measured value of 0.60 $\pm$ 0.02 $R_\odot$ (from 
equation~\ref{eqn:secradius}) is consistent with this value.

\begin{table*}
{\protect\small
\caption[System parameters]{System parameters for V347~Pup. The 
Monte Carlo results for corrected and uncorrected $K_R$ values are 
shown for comparison. The radial velocity of the white dwarf ($K_W$) has 
also been calculated from the secondary star parameters.}
\label{tab:params}
\begin{tabular*}{110mm}
{lr@{$\:\pm\:$}lr@{$\:\pm\:$}lr@{$\:\pm\:$}lr@{$\:\pm\:$}l} 
\\
\hline
\vspace{-3mm}\\
\multicolumn{1}{l}{Parameter} & \multicolumn{4}{c}{Non $K_R$--corrected} & 
\multicolumn{4}{c}{$K_R$--corrected} \\
\multicolumn{1}{l}{} & \multicolumn{8}{c}{------------------------------------------------------------------------------------} \\
\multicolumn{1}{c}{} &
\multicolumn{2}{c}{Measured} & 
\multicolumn{2}{c}{Monte Carlo} & 
\multicolumn{2}{c}{Measured} & 
\multicolumn{2}{c}{Monte Carlo} \\ 
\multicolumn{1}{l}{} &
\multicolumn{2}{c}{Value} & 
\multicolumn{2}{c}{Value} &
\multicolumn{2}{c}{Value} &
\multicolumn{2}{c}{Value} \\
\vspace{-3mm}\\
\hline
\vspace{-3mm}\\
$P_{orb}$ (d) & \multicolumn{2}{c}{0.231936060} & \multicolumn{2}{c}{} & 
\multicolumn{2}{c}{0.231936060} & \multicolumn{2}{c}{} \\ 
$K_R$ (km\,s$^{-1}$) & 216&5 & 215&5 & 198&5 & 198&5 \\ 
$v \sin i$ (km\,s$^{-1}$) & 130&5 & 131&5 & 130&5 & 131&5 \\
$\Delta\phi_{1/2}$ & 0.115&0.005 & 0.111&0.003 & 0.115&0.005 & 0.113&0.004 \\ 
$q$ & \multicolumn{2}{c}{} & 0.73&0.05 & \multicolumn{2}{c}{} & 0.83&0.05 \\ 
$i^\circ$ & \multicolumn{2}{c}{} & 85.0&2.1 & \multicolumn{2}{c}{} & 84.0&2.3 \\
$K_W$ (km\,s$^{-1}$) & \multicolumn{2}{c}{} & 158&9 & 
\multicolumn{2}{c}{} & 163&9 \\
$M_1/M_\odot$ & \multicolumn{2}{c}{} & 0.73&0.05 & 
\multicolumn{2}{c}{} & 0.63&0.04 \\
$M_2/M_\odot$ & \multicolumn{2}{c}{} & 0.54&0.06 & 
\multicolumn{2}{c}{} & 0.52&0.06 \\
$R_2/R_\odot$ & \multicolumn{2}{c}{} & 0.60&0.02 & 
\multicolumn{2}{c}{} & 0.60&0.02 \\
$a/R_\odot$ & \multicolumn{2}{c}{} & 1.72&0.04 & 
\multicolumn{2}{c}{} & 1.66&0.05 \\
d (pc) & 490&130 & \multicolumn{2}{c}{} & 
490&130 & \multicolumn{2}{c}{} \\
Spectral type & \multicolumn{2}{c}{M0.5 V} & \multicolumn{2}{c}{} & 
\multicolumn{2}{c}{M0.5 V} & \multicolumn{2}{c}{} \\
of secondary & \multicolumn{2}{c}{} & \multicolumn{2}{c}{} 
& \multicolumn{2}{c}{} & \multicolumn{2}{c}{} \\
$\Delta\phi$ & 0.110&0.005 & \multicolumn{2}{c}{} & 0.110&0.005 & 
\multicolumn{2}{c}{} \\
$R_D/R_1$ & \multicolumn{2}{c}{} & 0.72&0.08 & \multicolumn{2}{c}{} & 0.72&0.09 \\
\hline\\
\end{tabular*}
}
\end{table*}

\section{Discussion}
\label{sec:discussion}

\subsection{Spiral arms}
\label{sec:shocks}

Spiral-armed disc asymmetries are evident in the HeI and H$\beta$ Doppler
maps, confirming the findings of \scite{still98} in their H$\beta$ and 
H$\gamma$ maps. Similar spiral structures
have been observed in dwarf novae during outburst (e.g. IP~Peg, 
\pcite{steeghs97}; U Gem, \pcite{groot01}).  Tidally
driven spiral density waves can develop in accretion discs due to the
tidal torque of the mass donor star on the outer disc (\pcite{sawada86}, 
\pcite{blondin00}, \pcite{boffin01}).  Their detection in outburst only 
reflects the much stronger tidal effects on the accretion disc when it
increases in size and temperature during outburst, in which case a tidally
induced spiral structure is expected that closely matches the observed
structures (\pcite{armitage98}, \pcite{steeghs99}, \pcite{steeghs01}).  
In dwarf novae, these asymmetries decay as the system returns to quiescence, 
and the disc cools and shrinks. In order for a similar tidal response to be
responsible for the disc asymmetry in V347 Pup, its disc must be
large and comparable to the tidal radius. We calculate the tidal radius of the 
accretion disc to be 0.33$a$ using the 
pressureless disc models of \scite{paczynski77} and our new system parameters. 
The measured disc radius of $R_D/a$ = 0.28 $\pm$ 0.03 is comparable in size 
to the tidal radius, and therefore consistent with a tidal origin for the 
observed spiral structure.

Our observations show that the spiral structures are clearly visible in the 
HeI Doppler maps, but are either weak or non-existent in the Balmer and HeII 
maps. This is in contrast to dwarf novae in outburst, which typically show 
stronger spiral structures in the HeII and Balmer lines (e.g. 
\pcite{marsh00}; \pcite{morales04}). This could be a reflection of 
different densities and temperatures in NL discs compared to the discs of 
dwarf novae in outburst, or it could 
simply be due to a contrast effect where the relative contribution of the 
spiral structure is not as high in the HeII and Balmer maps due to the 
presence of low--velocity emission.

Note that the impact of such tidally--induced spiral arms on the
angular momentum transport has not been fully established. If they are
associated with hydrodynamical shocks, such as in the 
simulations of \scite{sawada86}, their contribution to the angular
momentum transport could be very significant. On the other hand, 
\scite{smak01} and \scite{ogilvie02} propose that these disc structures may 
reflect tidally thickened areas in the outer disc as it expands 
close to its tidal radius. Their enhanced emission is then caused by 
irradiation from the accreting white dwarf and regions close to it. 

The prospect of testing such basic disc physics with observations
warrants the study of these disc structures in more detail (see also
\pcite{morales04}). With V347 Pup, we have a target that appears to have 
a persistent disc asymmetry that is more accessible than the transient 
spiral structure observed in dwarf novae. 

\subsection{Mass transfer stability}
\label{masstrans}

The mass ratio of a CV is of great significance, as it 
governs the properties of mass transfer from the secondary to the white 
dwarf primary. This in turn governs the evolution and behaviour 
of the system. 

The secondary star responds to mass loss on two timescales. First, the 
star returns to hydrostatic equilibrium on the dynamical timescale, which is 
the  sound--crossing time of the region affected. Second, the star settles 
into a new thermal equilibrium configuration on a thermal timescale.

\begin{figure}
\begin{center}
\includegraphics[width=8.4cm,angle=-90]{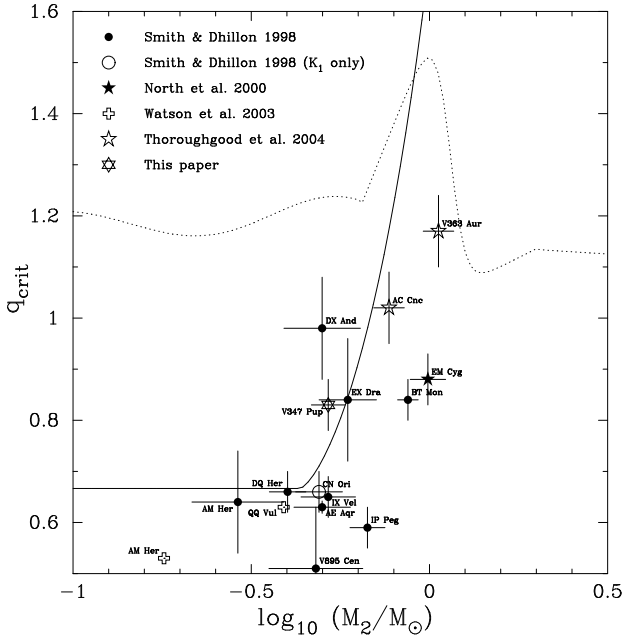}
\end{center}
\caption{\protect\small
Critical mass ratios for mass transfer stability. The dotted line represents 
the condition for thermal instability; the solid line represents the 
condition for dynamical instability (Politano 1996). 
Both curves assume the star is initially in thermal equilibrium. 
Mass ratios and secondary masses from the compilation of 
\protect\scite{smith98}, \protect\scite{north00}, \protect\scite{watson03}, 
and \protect\scite{thoroughgood04} are overplotted. 
The mass ratios and secondary star masses of V347~Pup 
determined in this paper are also plotted.}
\label{fig:qplot}
\end{figure}

The two timescales upon which the secondary responds to mass loss leads to
two types of mass transfer instability. If, upon mass loss, the dynamical
response of the secondary is to expand relative to the Roche lobe, mass
transfer is dynamically unstable and mass transfer proceeds on the
dynamical timescale. \scite{politano96} made an analytic fit 
to the models of \scite{hjellming89} to give the limit of dynamically stable 
mass loss, plotted as the solid line in Fig.~\ref{fig:qplot}. Dynamically 
stable mass transfer can occur if the CV lies below this line. 
This limit is important for low mass secondary stars ($M_2 < 0.5M_\odot$), as 
they have significant convective envelopes that tend to expand adiabatically 
in response to mass loss (\pcite{dekool92}). 

Thermally unstable mass transfer is possible if the dynamic response
of the star to mass loss is to shrink relative to its Roche lobe
(i.e. mass transfer is {\em dynamically} stable). 
This occurs at high donor masses ($M_2 > 0.8M_\odot$) when the 
star has a negligible convective envelope and its adiabatic response 
to mass loss is to shrink. (e.g. \pcite{dekool92}; \pcite{politano96}). 
Mass transfer then
initially breaks contact and the star begins to settle into its new thermal
equilibrium configuration. If the stars thermal equilibrium radius is
now bigger than the Roche lobe, mass transfer is again unstable, but
proceeds on the slower, thermal timescale. 
The limit of thermally--stable mass transfer can be 
found by differentiating the main--sequence mass--radius relationship 
given in \scite{politano96}.
Thermally--stable mass transfer can occur if the CV appears below the dotted 
line plotted in Fig.~\ref{fig:qplot}. 

The limit for dynamically stable mass transfer is important in the case of 
V347~Pup owing to the low secondary star mass. Fig.~\ref{fig:qplot} shows 
that the system is just consistent with the limit at the 1$\sigma$ level. 
The mass transfer stability limits, however, are true only for 
ZAMS stars, whereas the secondary stars in CVs are expected to have undergone 
some evolution. The loss of the outer envelope, 
for example, would result in a larger than normal helium to hydrogen ratio and 
affect the star's response to mass loss. For instance, DX~And, which lies 
outside the limit, has been shown to have an evolved companion (\pcite{drew93}). 

There is tentative evidence that the secondary star in V347~Pup is evolved by 
considering three pieces of evidence. 
First, V347~Pup falls outside the limit for dynamically 
stable mass transfer (although agrees at the 1$\sigma$ level). Second, 
the measured radius is at the upper limit for a main--sequence companion 
of the same mass (\pcite{smith98}). Third, the secondary star mass and 
spectral type measured for V347~Pup are closer to the evolved models of 
\scite{kolb01} than the ZAMS models.

\section{Conclusions}
\label{sec:conclusions}

\begin{enumerate}

\item{We have measured the radial and rotational velocities of the secondary 
star in V347~Pup in order to calculate the component masses and other 
system parameters. The secondary star radial velocity is affected by 
irradiation from the emission regions around the primary, which we correct 
for using a model. We find the component masses in V347~Pup to be 
$M_1$ = 0.63 $\pm$ 0.04 $M_\odot$ for the white dwarf primary and 
$M_2$ = 0.52 $\pm$ 0.06 $M_\odot$ for the M0.5V secondary star. 
V347~Pup shows many of the characteristics of the SW~Sex stars, 
exhibiting single--peaked emission lines, high--velocity S--wave 
components and phase--offsets in the radial velocity curves.}

\item{V347~Pup lies outside the theoretical limit for dynamically 
stable mass transfer in ZAMS stars, but is just consistent at the 
1$\sigma$ uncertainty level. This piece of evidence, together with a 
secondary star radius at the upper limit for a main--sequence 
companion of the same mass, suggests that the secondary star in 
V347~Pup may be evolved. Additionally, the secondary star mass and 
spectral type measured for V347~Pup are closer to the evolved models of 
\scite{kolb01} than the ZAMS models.}

\item{The presence of spiral arms in the accretion disc, first noted 
by \scite{still98}, has been confirmed. Consistent with this, we find that 
the measured accretion disc radius is close to the tidal radius computed from 
the pressureless disc models of \scite{paczynski77}. The persistent 
spiral arms seen in this bright novalike makes it an excellent candidate in which 
to study these features, rather than the transient spiral structures observed in 
dwarf novae.}

\end{enumerate}

\section*{\sc Acknowledgements}
TDT is supported by a PPARC studentship; CAW is supported by PPARC 
grant number PPA/G/S/2000/00598; SPL is supported by PPARC. 
DS acknowledges a Smithsonian Astrophysical Observatory Clay Fellowship. 

\bibliographystyle{mnras}
\bibliography{refs}

\end{document}